\documentclass{aa}  
\usepackage{graphicx}
\usepackage{txfonts}
\usepackage{bm} 
\usepackage{epstopdf}
\usepackage{multirow} 
\usepackage{rotating}
\usepackage{color}
\usepackage{xfrac}
\usepackage{ulem} 
\usepackage{natbib}
\usepackage{array} 
\usepackage{subcaption} 
\usepackage[colorlinks=true,citecolor=blue]{hyperref}

\newcolumntype{M}[1]{>{\centering\arraybackslash}m{#1}}
\newcolumntype{C}[1]{>{\centering\arraybackslash}m{#1}}
\newcolumntype{R}[1]{>{\raggedleft\arraybackslash}m{#1}}
\newcommand{\minitab}[2][l]{\begin{tabular}{#1}#2\end{tabular}}

\DeclareMathOperator{\rad}{rad}

\begin{document} 
   \title{Laboratory validation of the dual-zone phase mask coronagraph in broadband light at the high-contrast imaging THD-testbed}
   \titlerunning{Lab. validation of the DZPM coronagraph} 
   \authorrunning{J.Delorme}

   \author{J. R. Delorme\inst{1}
          \and
          M. N'Diaye\inst{2}
          \and          
          R. Galicher\inst{1}
          \and
          K. Dohlen\inst{3} 
          \and          
          P. Baudoz\inst{1}
          \and\\  
          A. Caillat\inst{3}          
          \and
          G. Rousset\inst{1}
          \and
          R. Soummer\inst{2}
          \and
          O. Dupuis\inst{1}
          }

   \institute{LESIA, Observatoire de Paris, CNRS, University Pierre et Marie Curie Paris 6 and University Denis Diderot Paris 7,
             \\5 place Jules Janssen, 92195 Meudon, France
             \and
            Space Telescope Science Institute, 3700 San Martin Drive, 21218 Baltimore MD, USA
             \and
             Aix Marseille Université, CNRS, Laboratoire d'Astrophysique de Marseille (LAM), UMR 7326, 13388 Marseille, France\\
             \email{jacques-robert.delorme@obspm.fr}
             }

   \date{Received 23 March 2016;}
 
  \abstract
   {Specific high contrast imaging instruments are mandatory to characterize circumstellar disks and exoplanets around nearby stars. Coronagraphs are commonly used in these facilities to reject the diffracted light of an observed star and enable the direct imaging and spectroscopy of its circumstellar environment. One important property of the coronagraph is to be able to work in broadband light.}
   {Among several proposed coronagraphs, the dual-zone phase mask coronagraph is a promising solution for starlight rejection in broadband light. In this paper, we perform the first validation of this concept in laboratory.}
   {First, we recall the principle of the dual-zone phase mask coronagraph. Then, we describe the high-contrast imaging THD testbed, the manufacturing of the components and the quality-control procedures. Finally, we study the sensitivity of our coronagraph to low-order aberrations (inner working angle and defocus) and estimate its contrast performance. Our experimental broadband light results are compared with numerical simulations to check agreement with the performance predictions.}
   {With the manufactured prototype and using a dark hole technique based on the self-coherent camera, we obtain contrast levels down to $2\,10^{-8}$ between 5 and 17$\,\lambda_0/D$ in monochromatic light ($640$\,nm). We also reach contrast levels of $4\,10^{-8}$ between 7 and 17$\,\lambda_0/D$ in broadband ($\lambda_0=675$\,nm, $\Delta\lambda=250$\ and $\Delta\lambda / \lambda_0 = 40$\%), which demonstrates the excellent chromatic performance of the dual-zone phase mask coronagraph.}
   {The performance reached by the dual-zone phase mask coronagraph is promising for future high-contrast imaging instruments that aim at detecting and spectrally characterizing old or light gaseous planets.}


   \maketitle
%

\section{Introduction}\label{Sec_Introduction}
During the past few years, high-contrast imaging instruments allowed the discovery and the spectral characterization of some objects in the outer part of exoplanetary systems. However, such detections are very challenging

During the past few years, high-contrast imaging instruments allowed the discovery and the spectral characterization of some objects in the outer part of extrasolar systems. However, imaging circumstellar disks and exoplanets is very challenging as these objects are $10^{3}$ to $10^{10}$ times fainter than their host stars in the visible and near-infrared domains with angular separations smaller than 1''\,\citep{Seager2010}. 

Starlight rejection techniques are required to overcome these contrast ratios and observe circumstellar environments. In this context, coronagraphy is a powerful technique that were adopted in the current high-contrast imaging instruments, such as SPHERE/VLT \citep{Beuzit2008}, GPI/Gemini \citep{Macintosh2008}, SCExAO/Subaru \citep{Jovanovic2014} and P1640/Palomar \citep{Hinkley2011}. Thanks to their coronanagraphs these instruments can study , in optimal observation conditions, young and massive planets, and thus warm and bright at infrared wavelength that are $10^{-7}$ fainter than their hosting star. The presence of advanced coronagraphs is already envisioned for the forthcoming extremely large telescopes (ELTs) on the ground and the future large missions in space \citep[e.g., WFIRST,  HDST][]{Spergel2015,Dalcanton2015} to obtain contrast levels down to $10^{-10}$ to study cold gaseous or massive rocky planets. Coronagraphs need highly achromatic capabilities to enable the spectral analysis of disk grains and exoplanet atmospheres with reasonable exposure times. Several concepts have been proposed to observe substellar mass companions over wide spectral bands \citep[e.g.][]{Guyon2005,Mawet2005,Trauger2011,Galicher2011}. Among them, the dual-zone phase mask (DZPM) coronagraph is a promising solution that combines a pupil apodizer and a focal dual-zone phase mask to remove starlight in broadband light \citep{Soummer2003b}. 

To validate the multireference self-coherent camera \cite[MRSCC, ][]{Delorme2016_MRSCC}, which is a focal plane wavefront sensor working in wide-band, we manufactured a DZPM coronagraph for the \textit{tr\`es haute dynamique} \cite[THD, ][]{Galicher2014} testbed. We choose the DZPM because it is attractive in term of contrast level ($\approx10^{-8}$) over a wide-band (up to 40\%) and it is easy to implement. Although the concept itself had never before been demonstrated, the technologies required for its manufacture had been demonstrated, both for the apodizer \citep{Martinez2009,Sivaramakrishnan2009} and for the phase mask \citep{N'Diaye2010}.

In this paper, we study a prototype of this device in laboratory. In Sect.\,\ref{Sec_DZPM} we recall the coronagraph principle. In Sect.\,\ref{Sec_Setup}, we present the THD testbed, detail the design of the adopted DZPM coronagraph for this testbed, and describe the manufacturing procedures for the apodizer and phase mask prototypes. In Sect.\,\ref{Sec_Lab}, we present the results of our experiment with the determination of the inner working angle (IWA) and the extinction of the DZPM, the analysis of its sensitivity to defocus, and the presentation of the contrast performance of our prototype for different spectral bandwidth up to 40\%.

\section{DZPM principle}
\label{Sec_DZPM}
\begin{figure}[h!]
        \centering
        \begin{subfigure}[b]{0.48\textwidth}
        \includegraphics[width = \textwidth]{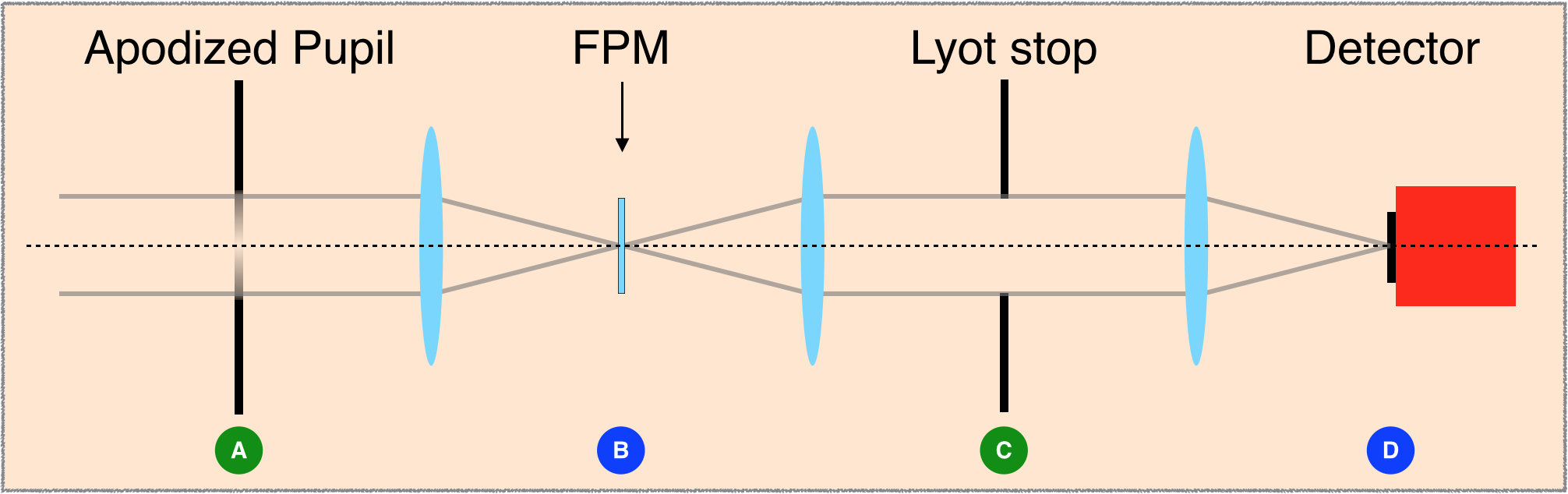}
        \end{subfigure}
        \caption{Scheme of the coronagraphic layout.}
        \label{layman}%
\end{figure}

A phase mask coronagraph is a Lyot-style concept which involves four successive planes (A, B, C, D) represented by Fig. \ref{layman}. It combines an entrance pupil with apodization in plane A, a downstream phase focal plane mask (FPM) in plane B, a Lyot stop in the relayed pupil plane C, and a detector in the final focal plane D in which the coronagraphic image of an unresolved, on-axis star is formed. The design is such that an optical Fourier transform operation relates the complex amplitudes of the electric field in two successive planes. We recall the principle of concepts using circularly axi-symmetrical phase FPM, first with the Roddier \& Roddier phase mask (RRPM) and then with a DZPM. We refer the reader to \citet{Soummer2003a,Soummer2003b} and \citet{N'Diaye2012a} for a complete description of the mathematical formalism. 

The RRPM uses a single circular phase disk that introduces a $\pi$ phase shift with respect to the rest of the mask in the image plane B. The disk diameter is close to that of the Airy disk (1.06\,$\lambda_{opt}/D$ for a circular aperture of diameter $D$ and design wavelength $\lambda_{opt}$). For a monochromatic source emitting at $\lambda_0$, the mask creates a destructive interference within the relayed pupil, rejecting most of the stellar light outside the Lyot stop. Adding an adequate apodization in the entrance pupil plane A results in the cancellation of the electric field over the relayed pupil in plane C, and therefore leads to a perfect starlight suppression in the focal plane D  \citep{BaudozPHD,Guyon2000,Soummer2003a}. Unfortunately, the performance of the corresponding system strongly decreases in broadband observation. This loss of coronagraphic performance comes from chromatism effects related to the inherent properties of the RRPM. 

\begin{figure}[h!]
        \centering
        \begin{subfigure}[b]{0.48\textwidth}
        \includegraphics[width = \textwidth]{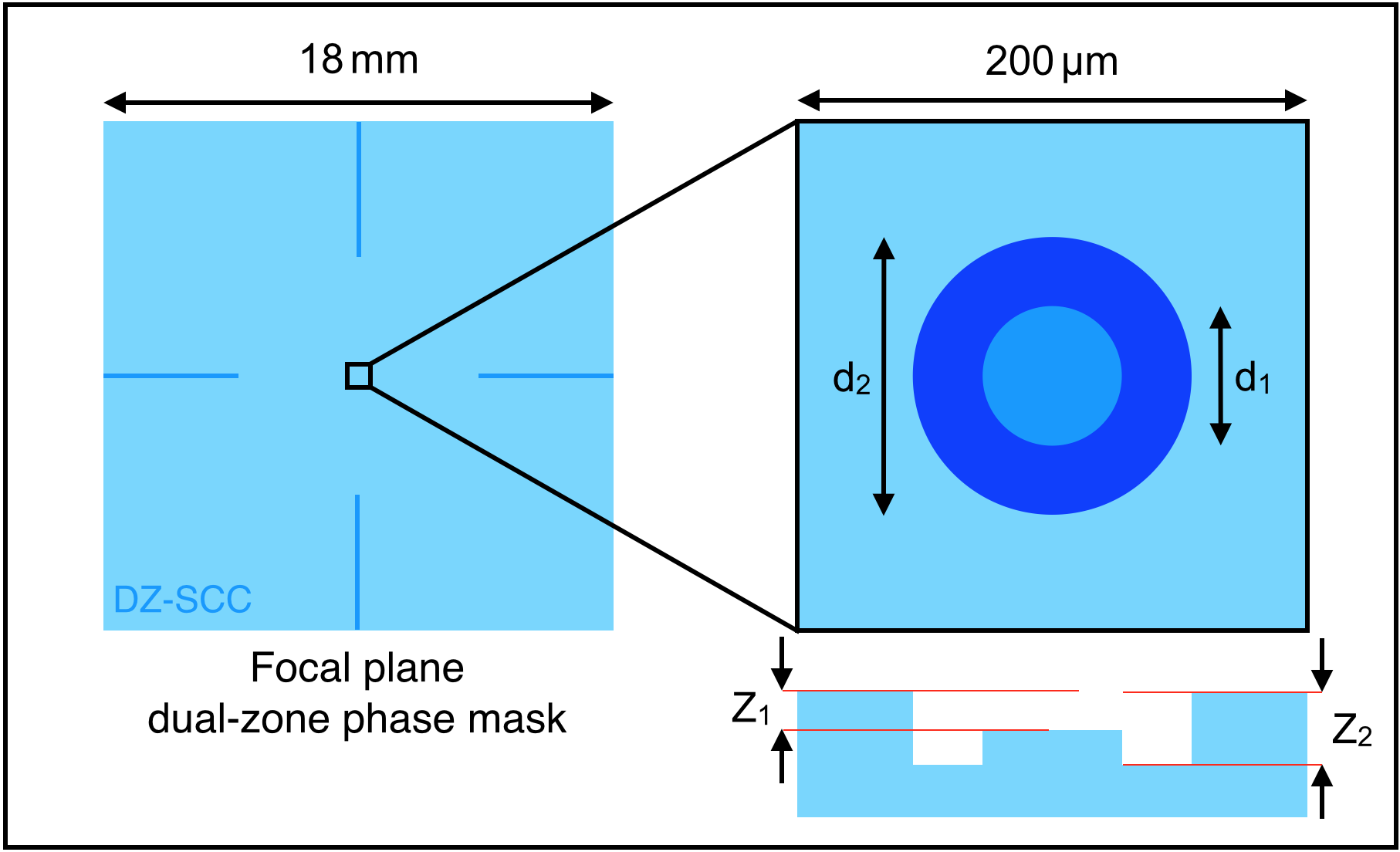}
        \end{subfigure}
        \caption{Substract layout and mask profile.}
        \label{Fig_Phase_Mask_diagram}%
\end{figure}

To overcome the chromaticity issue, \citet{Soummer2003b} improved the concept by replacing the simple phase mask in the Roddier coronagraph with a DZPM. This phase mask is designed as a circular phase disk of diameter $d_1$ surrounded by an annular phase ring of diameter $d_2$ where $d_1$ and $d_2$ are of the order of the Airy disk (see Fig.\,\ref{Fig_Phase_Mask_diagram}). The DZPM introduces different phase shifts $\varphi_1$ and $\varphi_2$ on the incoming wavefront with respect to the rest of the phase mask in the focal plane B. These phase steps are induced by the inner and outer part of the mask and expressed in optical path differences $OPD_1$ and $OPD_2$. At a given central wavelength $\lambda_{opt}$, the mask creates a destructive interference within the relayed pupil, rejecting the stellar light outside the Lyot stop. An adequate choice of the DZPM parameters allows reaching destructive interferences inside the pupil over a wide range of wavelengths. One of the main parameters is the entrance pupil apodizer which transmission follows a radially symmetric fourth order polynomial function to improve the starlight rejection  \citep{Soummer2003b}. In the case of a clear aperture and with normalized radial coordinate $\rho = r/R$ where r is the radial coordinate and R is the pupil radius, the amplitude apodization $\Phi_a$ writes as
\begin{equation}
\Phi_a (r)=1 + \omega_{1}\:\rho^{2} + \omega_{2} \:\rho^{4}\,,
\label{eq:apodizer}
\end{equation}
with the polynomial transmission function coefficients $\omega_{1}$ and $\omega_{2}$. Further improvement is achieved by adding a slight defocus of the coronagraph mask, which is equivalent to allocate a complex term in the apodization. The phase apodization $\Phi_w$ at the wavelength $\lambda$ is then expressed as
\begin{equation}
\Phi_w (r)= \exp (2i\pi \: \beta \: \rho^{2} \: \lambda_{opt} / \lambda)\,,
\end{equation}
with its associated parameter $\beta$ which longitudinal defocus $\Delta z$ is related to by 
\begin{equation}
\Delta z=2 \beta F^2 \lambda\,,
\end{equation}
where $F$ denotes the focal ratio of the optical system. The DZPM coronagraph involves seven parameters ($\omega_{1}$, $\omega_{2}$, $\beta$, $d_1$, $d_2$, $OPD_1$, $OPD_2$) and a careful choice of their values provides a design with a contrast better than $10^{-6}$ over 20\% bandpass at an angular separation of 2\,$\lambda_{opt}/D$ from the star \citep{Soummer2003b,N'Diaye2012a}.   

Over the past few years, DZPM prototypes were manufactured for tests in laboratory \citep{N'Diaye2012b} but so far, they have never been introduced in a testbed with sufficiently high wavefront control performance in the low and mid-spatial frequencies to confirm the predicted performance. We decided to probe the efficiency of the DZPM coronagraph (apodizer and phase mask) by testing a recently manufactured prototype at the THD-testbed at Paris Observatory.  

\section{Experimental setup}
\label{Sec_Setup}

In this section, we briefly recall the goals and the main components of the optical THD-testbed. Then, we present the DZPM coronagraph that was designed for this facility. Finally, we detail the manufacturing procedure and we present the metrology of the manufactured apodizer and phase mask.

\subsection{THD-testbed}
\label{SubSec_THD}

\begin{figure*}
        \centering
        \begin{subfigure}[b]{0.70\textwidth}
        \includegraphics[angle = 0,width = \textwidth]{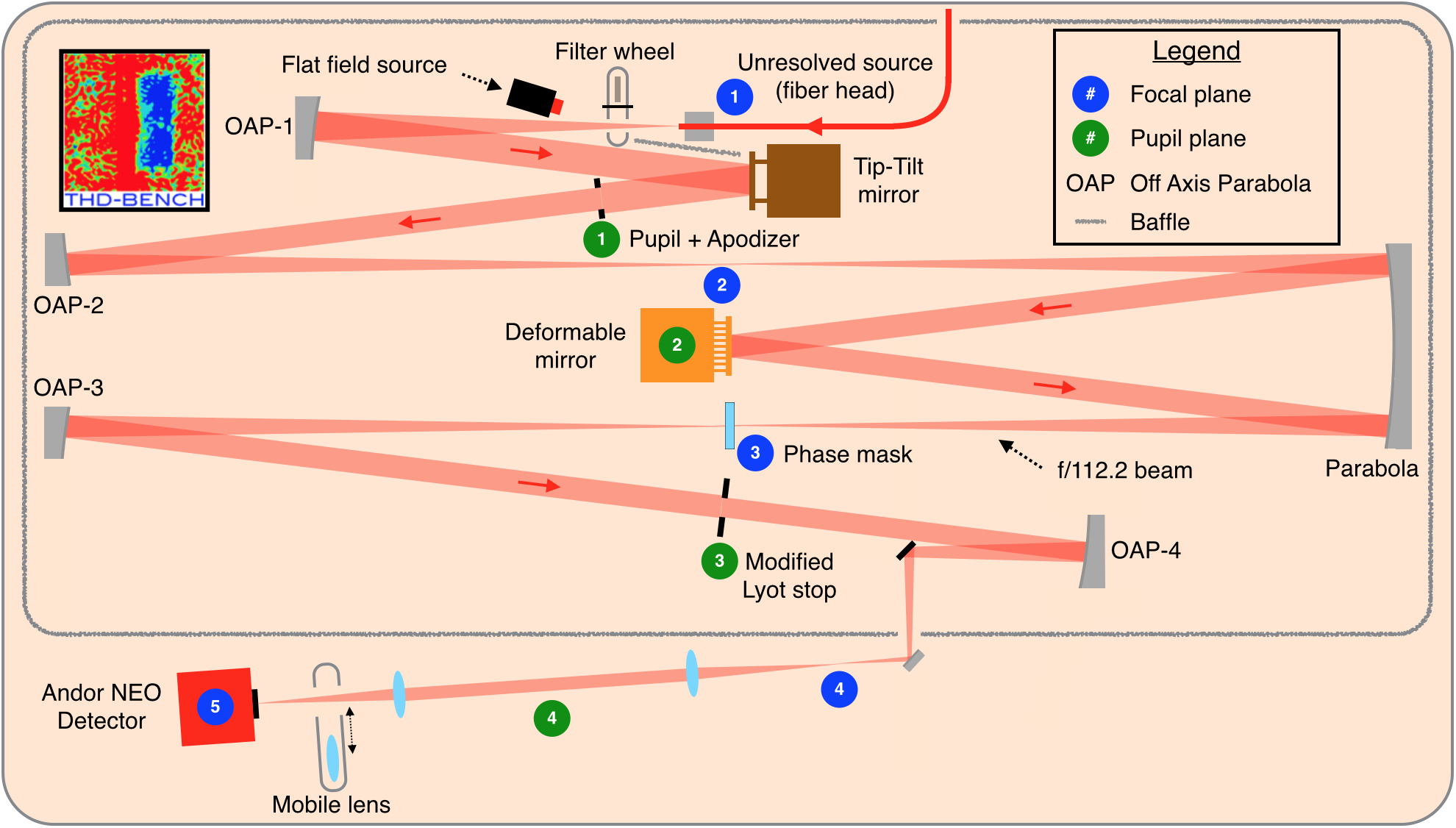}     
        \end{subfigure}
        \caption{Optical layout of the THD testbed.}
        \label{Fig_THD}%
\end{figure*}

The THD-testbed objective is the comparison of high contrast imaging techniques under the same conditions to prepare the design of future instruments dedicated to direct imaging of exoplanets \citep{Galicher2014}. During the experiments that are presented in this paper, a Self-Coherent Camera \citep[SCC,][]{Baudoz2006,Galicher2008,Galicher2010} is used as a focal plane wavefront sensor to minimize the speckle intensity in the final image by commanding a deformable mirror (DM) which is set upstream of the coronagraphic focal plane mask. A complete description of the THD-testbed (Fig.\,\ref{Fig_THD}) is given in \cite{Mas2010}. In this paper we only recall the main components.

\begin{itemize}
\item The star is simulated by a monomode optical fiber that can be fed by two light sources: a quasi-monochromatic laser diode (central wavelength $\lambda_{0} =$ 637 nm, spectral bandwidth $\Delta\lambda < 1$ nm); or a Fianium supercontinuum source with calibrated spectral filters (Tab.\,\ref{Tab_light_sources}).
\item A tip-tilt mirror is used to center the beam on the coronagraphic mask with pointing precision better than $1.5\:10^{-3} \lambda_{0}/D$.
\item The apodizer is set in the entrance circular unobscured pupil (diameter \O$_{p} $ =  8.1 mm) just after the tip-tilt mirror.
\item The DM is a 32$\times$32 actuators Boston Micromachines Corporation device. It is set in a pupil plane upstream the FPM. We used 27 actuators across the pupil diameter.
\item A modified Lyot stop is used to implement the SCC. The diameter of the classical Lyot stop aperture is $D_{L} $ = 8 mm (98.7\% filtering). The SCC reference hole is $\gamma$ = 26.6 times smaller and is at a distance $\xi$ = 1.76\,$D_{L}$ from the Lyot stop center.
\item An Andor NEO detector is used.
Images are 400$\times$400 pixels and the resolution element ($\lambda_{0} f/D_{L}$) is sampled with 6.25 pixels.
\item An additional lens on a motorized stage enables imaging of the light distribution in the pupil plane.
\end{itemize}

We classify light sources in four categories (Tab.\,\ref{Tab_light_sources}): monochromatic, 10 nm bandwidth, large bandwidth (< 100 nm) and very large bandwidth (> 200 nm). The monochromatic light source is the laser diode with no filter whereas the other light sources are produced using a combination of the supercontinuum source and one spectral filter or two (low-pass and high-pass). For all of our tests during the experiment, we extract 1\% of the injected light on the testbed to measure and record the source spectrum. These spectra have been used in the numerical simulations presented in this paper to reproduce the experimental conditions.
\begin{table}[h!]
\caption{Light sources used in this paper.} 
\centering
\begin{tabular}{|C{2.2cm}|C{1.8cm}|C{1.8cm}|C{1.7cm}|}
\hline
Light source category & Light source name & Central wavelength $\lambda_{0}$ (nm) & Bandwidth $\Delta\lambda$ (nm)\\
\hline
\hline
Monochromatic & Laser  & 637 & $<$1 \\
\hline
\hline
              & 620-10 & 620 & 10 \\
\cline{2-4}
 10 nm        & 647-10 & 647 & 10 \\
\cline{2-4}
bandwidth     & 660-10 & 660 & 10 \\
\cline{2-4}
              & 680-10 & 680 & 10\\
\hline
\hline
Large         & 652-30 & 652 & 30 \\
\cline{2-4}
bandwidth     & 641-80 & 641 & 80 \\
\hline
\hline
\multirow{3}*{\minitab[c]{Very large \\ bandwidth}} & 650-200 & 650 & 200 \\
\cline{2-4}
& 675-250 & 675 & 250 \\
\cline{2-4}
& 700-300 & 700 & 300 \\
\hline
\end{tabular}
\label{Tab_light_sources}
\end{table}

\subsection{DZPM coronagraph}
\label{SubSec_Optimisation}

\subsubsection{Design}
\label{SubSubSec_Design}

The DZPM coronagraph aims at rejecting the light of an on-axis star over a wide spectral band. The prototype optimized for the THD-testbed is designed to work over a bandwidth $\Delta\lambda = 133$\,nm centered at $\lambda_{opt} = 665$\,nm ($\Delta\lambda/\lambda_{opt} = 20$\%). Following \citep{N'Diaye2012a}, we adjust the DZPM parameters to reach the best contrast in the coronagraphic image over an annulus ranging from 2 to 10\,$\lambda_{opt}/D$ from the optical axis and over the spectral band of interest. For each set of parameters, we calculate the average intensity in the annulus at five uniformly distributed wavelengths. Then, we quadratically sum the five intensities. Eventually, we use a least-squares method to find the DZPM parameters that minimize the quadratic sum. Over the spectral bandpass mentioned above, the prototype theoretically provides a contrast better than $10^{-6}$ between 2 and 10\,$\lambda_{opt}/D$ from the star and $3\,10^{-8}$ at separations larger than 6\,$\lambda_{opt}/D$. We also expect it to reject the stellar light down to $5\,10^{-6}$ between 2 and 10\,$\lambda_{opt}/D$ and $2\,10^{-7}$ at separations larger than 6\,$\lambda_{opt}/D$ over 250\,nm.

\begin{figure}[h!]
        \centering
        \begin{subfigure}[b]{0.48\textwidth}
        \includegraphics[trim = 6mm 12mm 10mm 16mm, angle=90, clip, width = \textwidth]{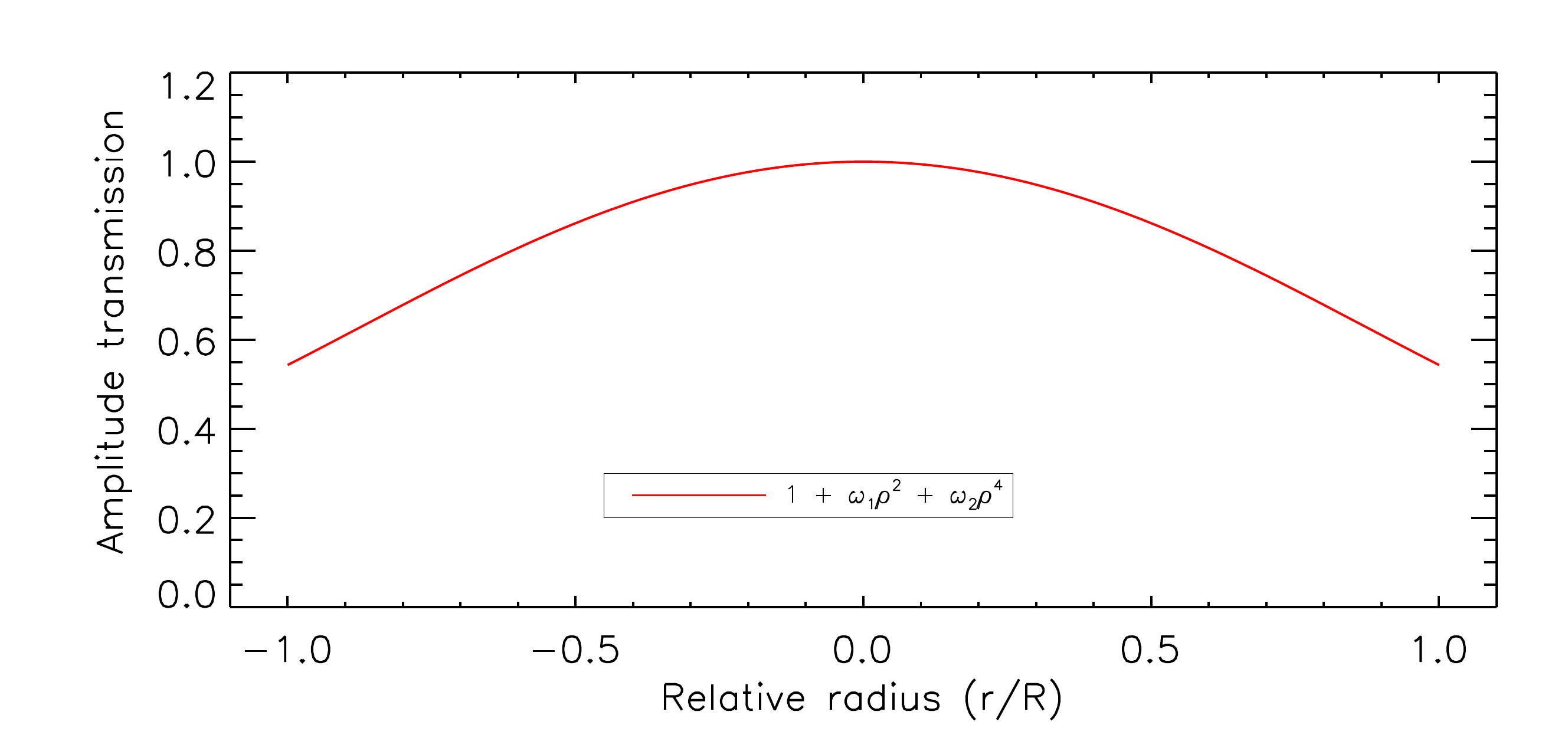} \\         
        \end{subfigure}
        \caption{Amplitude profile of the designed apodizer with $\omega_{1} = -0.584907$ and $\omega_{2} = 0.128201$.}
        \label{Fig_Ampl_apod}%
\end{figure}

The FPM parameters specified are given in Tab.\,\ref{Tab_Phase_mask} and the amplitude transmission profile of the apodizer, defined by Eq.\,\ref{eq:apodizer} with $\omega_{1} = -0.584907$ and $\omega_{2} = 0.128201$. The radius of the pupil is R = 4.05\,mm. The profile is plotted in Fig.\,\ref{Fig_Ampl_apod}. The throughput in intensity of the apodizer is 56\%  and the FWHM of the apodized PSF is 1.17\,$\lambda/D$.

In Sect.\,\ref{SubSec_Apodizer} and \ref{SubSec_Phase_Mask}, we describe the manufacturing procedures and our verifications of specification for both the apodizer and the phase mask.

\subsubsection{Apodizer characterization}
\label{SubSec_Apodizer}

\begin{figure}[h!]
        \centering
        \includegraphics[ width = .24\textwidth]{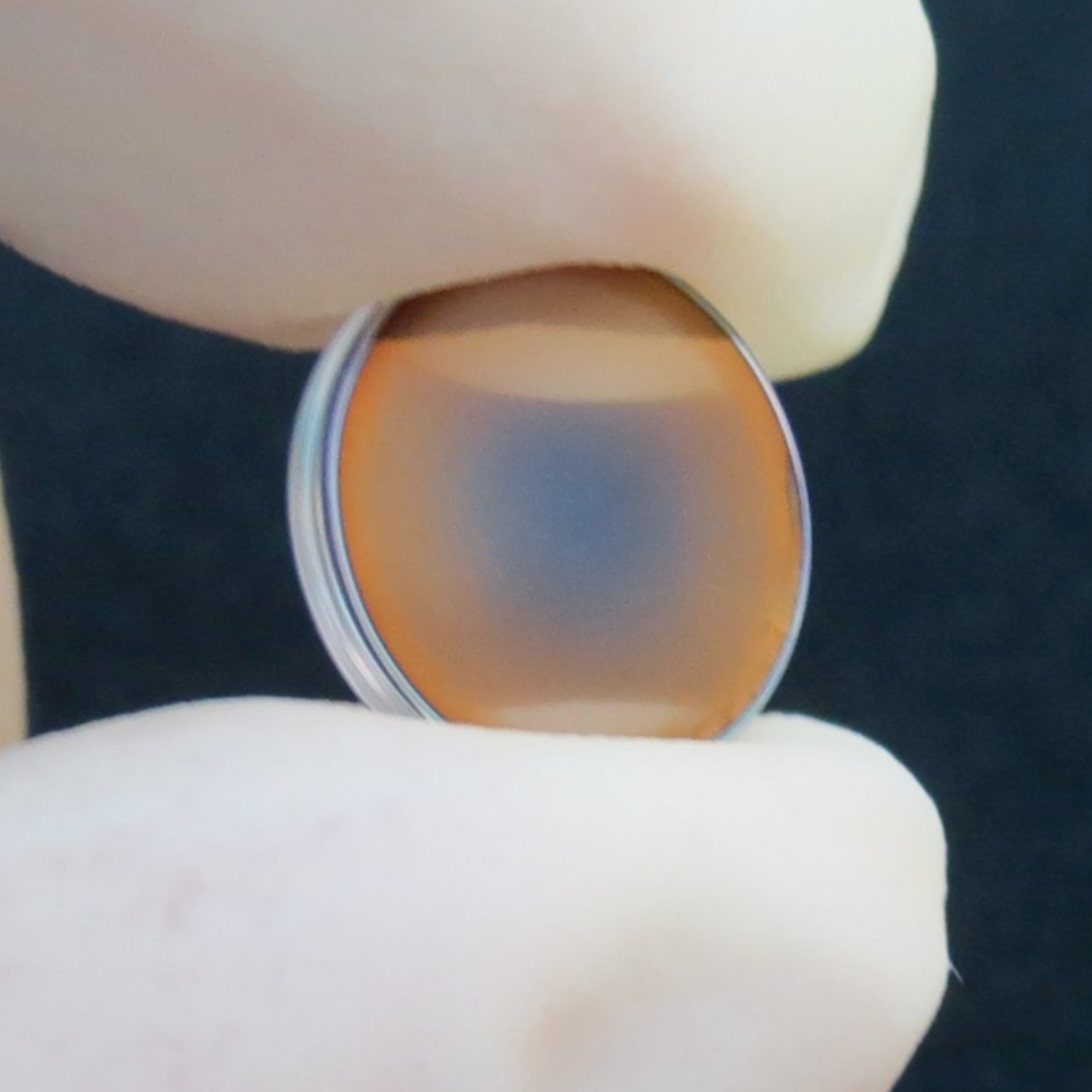}
        \includegraphics[width = .24\textwidth]{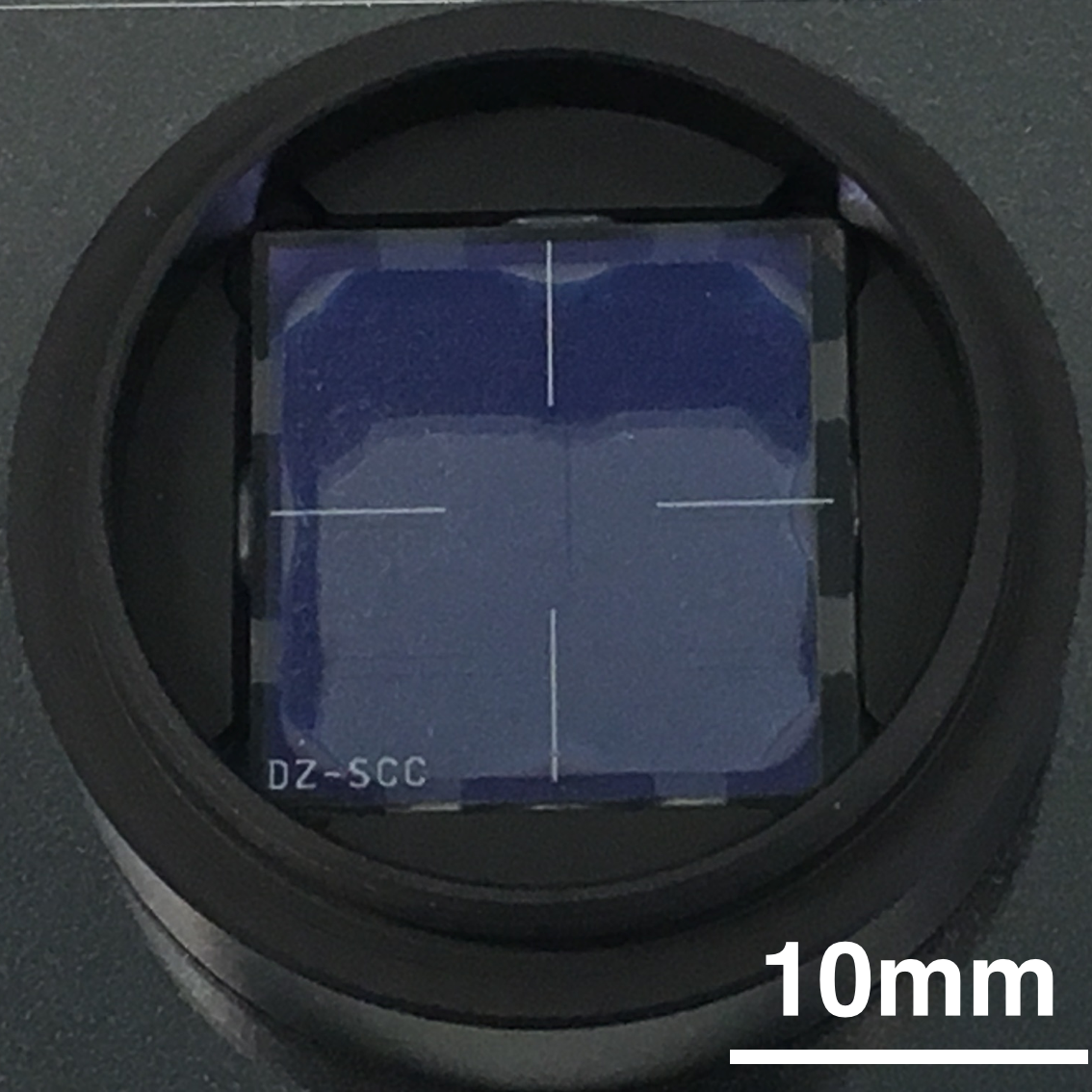}
        \caption{Picture of: the apodizer (left) and the phase mask in its assembly (right).}
        \label{Fig_Apodizer_Picture}
\end{figure}

The apodizer is manufactured by Aktiwave LLC (Fig.\,\ref{Fig_Apodizer_Picture}) using a halftone process \citep{Dorrer2007,Martinez2009,Vigan2015_apod} to reach the required transmission function through the use of an array of microdots (5$\mu$m by 5$\mu$m) with a customized spatial distribution over the 8.1 mm diameter. The same manufacturing process has been used for the SPHERE \citep{Guerri2011,Carbillet2011,Soummer2005} and GPI \citep{Sivaramakrishnan2010} apodizers.

In this section we present two tests done on the THD-testbed: the first to ensure that the apodizer amplitude profile is within the specifications; the second to measure its chromatic variation. In both cases, we record two images of the light intensity distribution in pupil plane: with and without the apodizer in the beam. Then, we divide the former by the latter and take the square root to obtain the normalized apodizer amplitude map. Finally, we derive the azimuthal average profile of this map to obtain the apodizer amplitude profile.
\begin{figure}[h!]
        \centering
        \includegraphics[width = .24\textwidth]{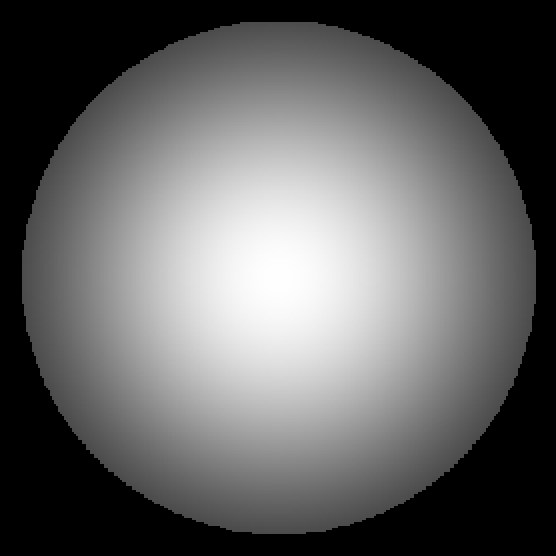}
        \includegraphics[width = .24\textwidth]{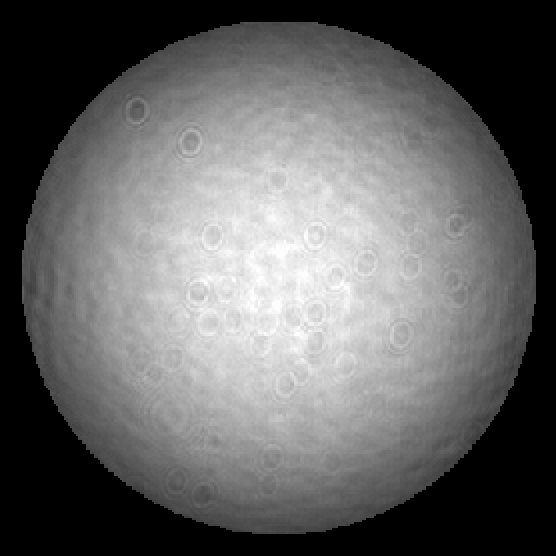}
        \caption{Apodizer amplitude map: specification (left) and laser measurement (right).}
        \label{Fig_Apod_im}
\end{figure}

Figure\,\ref{Fig_Apod_im} shows the expected apodizer amplitude map (left) and a laboratory measurement (right) obtained in monochromatic light. Dusts are visible in the experimental image as well as speckles due to the coherence of the laser.
\begin{figure}[h!]
        \centering
        \includegraphics[angle=90, width = .48\textwidth]{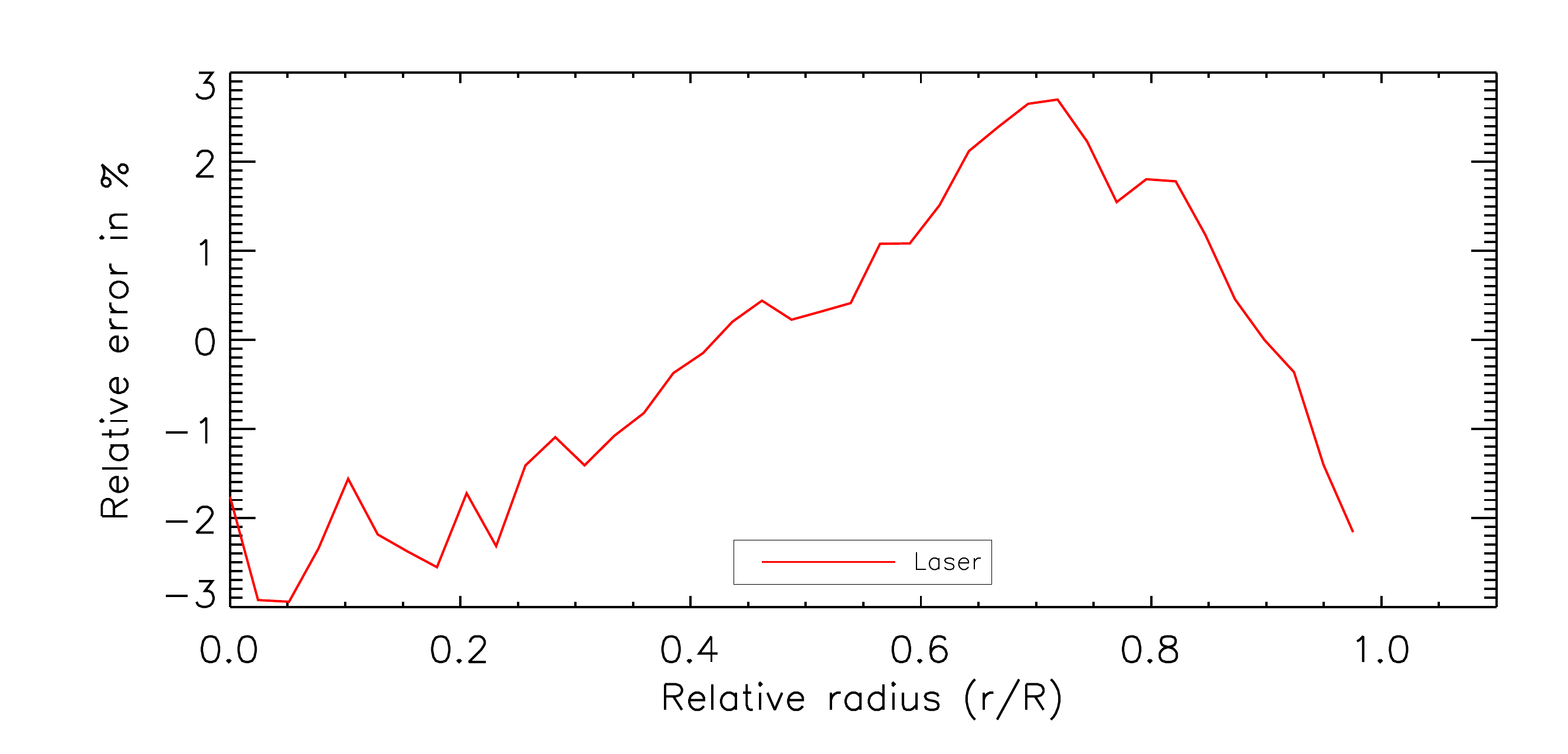}
        \caption{Apodizer amplitude profile: relative difference between laser measurement and specified function (Fig. \ref{Fig_Ampl_apod}).}
        \label{Fig_Laser_vs_Theorique}%
\end{figure}

Figure\,\ref{Fig_Laser_vs_Theorique} shows the relative difference between the apodizer amplitude profile measured in the laser image (Fig.\,\ref{Fig_Apod_im}, right) and the specifications provided to the manufacturer (Fig.\,\ref{Fig_Ampl_apod}). We find a difference smaller than $3\%$, which includes small amplitude errors of the THD-testbed and the effects of dusts.

\begin{figure}[h!]
        \centering
        \includegraphics[angle=90,clip, width = .48\textwidth]{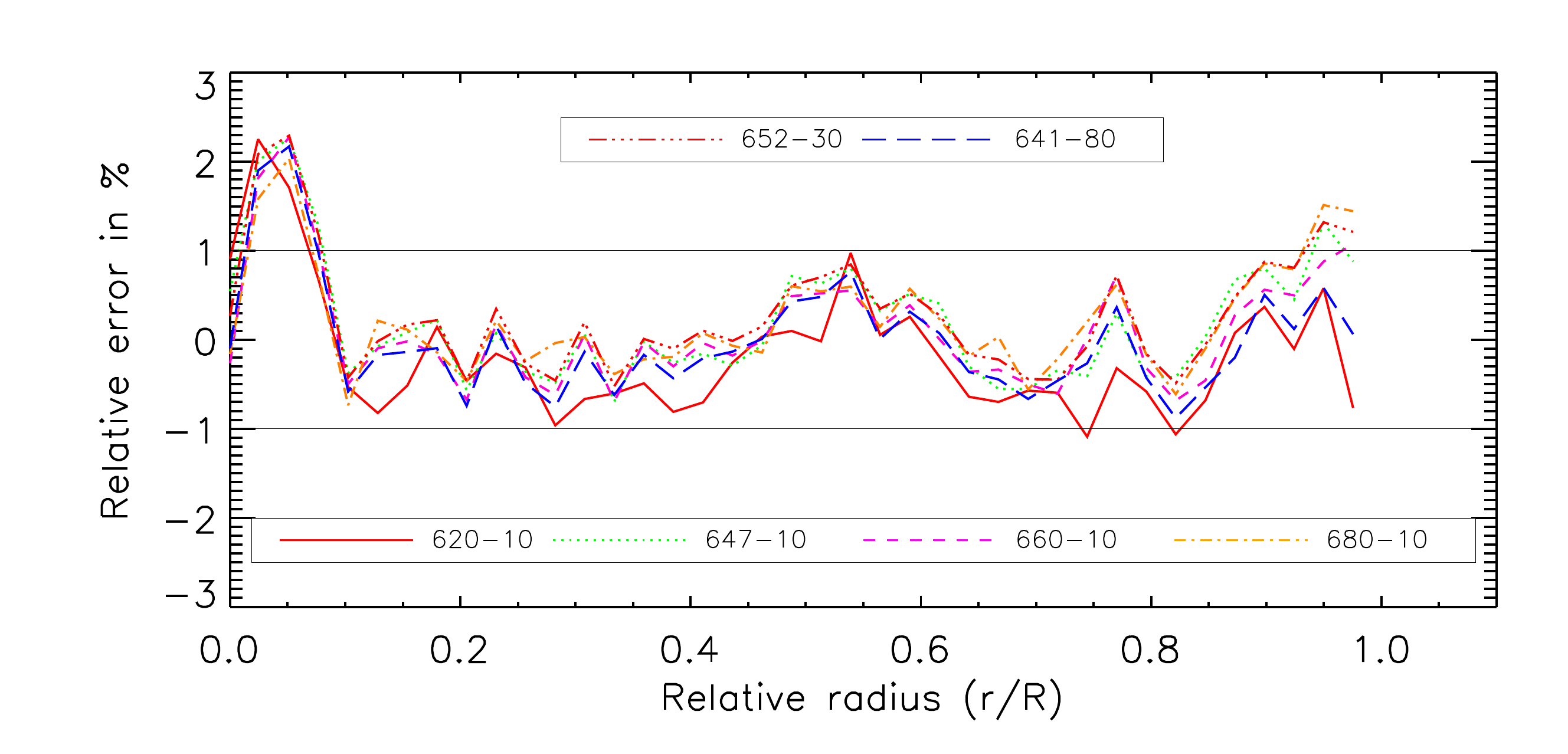}
        \caption{Apodizer amplitude profile: relative error between the laser and the other filters.}
        \label{Fig_Sources_vs_laser}%
\end{figure}

To probe the chromaticity of the apodizer, we measured the apodizer amplitude profile for all the filters in Tab.\,\ref{Tab_light_sources} with bandwidth smaller than 100\,nm. Figure\,\ref{Fig_Sources_vs_laser} shows the relative difference between the laboratory laser apodization profile (derived from Fig.\,\ref{Fig_Apod_im}, right) and those obtained for each filter: 10\,nm bandwidth filters centered on 620 nm (red solid curve), 647 nm (dotted green curve), 660 nm (dashed blue curve), 680 nm (dashed dotted orange curve), 30\,nm bandwidth filter (dot red curve) and 80\,nm bandwidth filter (dashed blue curve). For any light sources and for any radius, the relative difference is always smaller than $2.5\%$ showing good achromaticity of our component between 600 and 685\,nm.

\subsubsection{Phase mask characterization}
\label{SubSec_Phase_Mask}

The phase mask consists of two concentric cylinders shapes machined into the front face of a fused silica substrate by masking-photolithography and reactive ion-etching-processes, see Fig.\,\ref{Fig_Phase_Mask_diagram}. This subtractive process, already experimented and optimized in the context of single zone patterns such as the Roddier coronagraph \citep{N'Diaye2010,N'Diaye2011} and Zernike wavefront sensors \citep{Dohlen2006,N'Diaye2014b} have been found superior to the more classical additive process where SiO$_2$ is deposited onto a fused silica substrate \citep{Guyon1999}. Indeed, while the reactive ion-etching-process offers extremely steep edges and precisely defined phase steps, it is also monolithic, avoiding any interfaces between materials giving rise to spurious interference effects. The phase mask (Fig.\,\ref{Fig_Apodizer_Picture}) is manufactured by the SILIOS Technologies company. In the first step, a circular hole with diameter $d_{2}$ and depth $Z_{1}$ is generated, while in the second step, an annular hole with a depth $Z_{2}-Z_{1}$ is generated by masking the center of the first hole on the diameter $d_{1}$. Its outer diameter is identical to that of the original hole ($d_{2}$). Besides achieving good control of the depth of each pattern, the most critical step is the alignment of the two patterns. Any misalignment gives rise to a ridge along the edge of the outer diameter. For each manufacturing step, the mask shape is transferred into photo resist by UV insulation and chemical revelation of the resin layer, leaving the surface to be machined naked. Reactive ion etching is then applied by exposing the surface to SF$_6$ plasma.
\begin{figure}[h!]
\centering
\resizebox{\hsize}{!}{\includegraphics{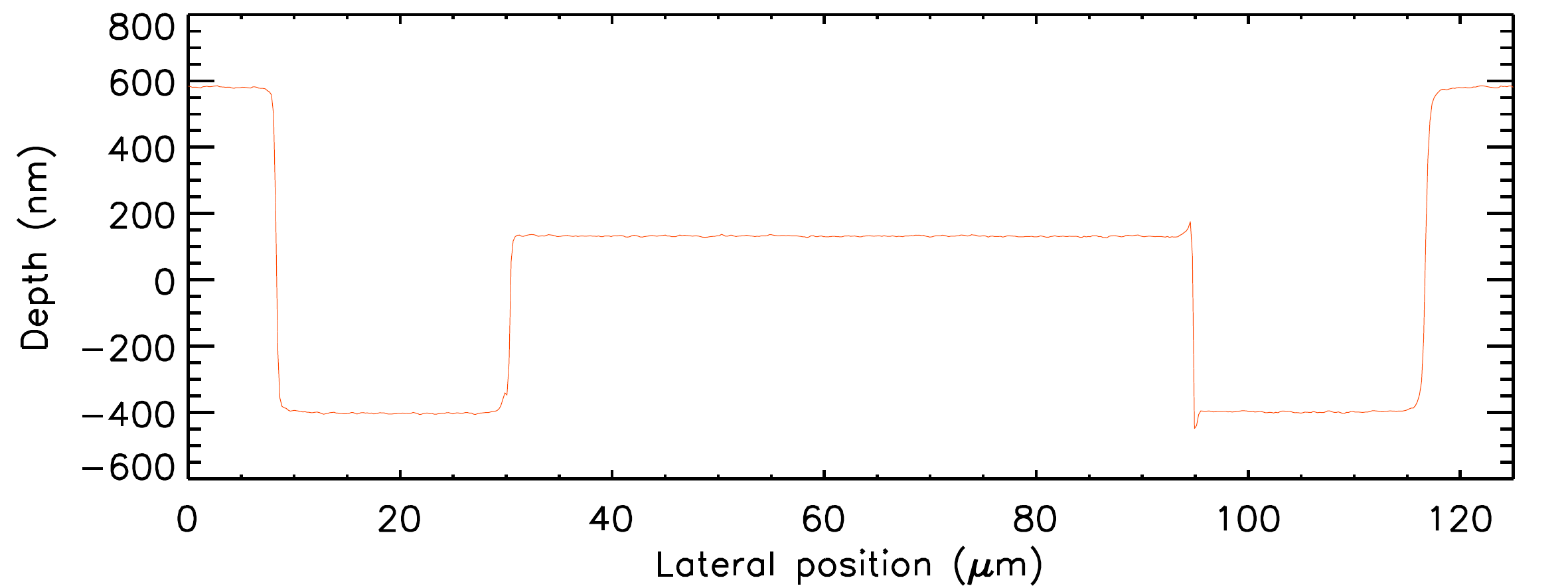}}
\resizebox{\hsize}{!}{\includegraphics{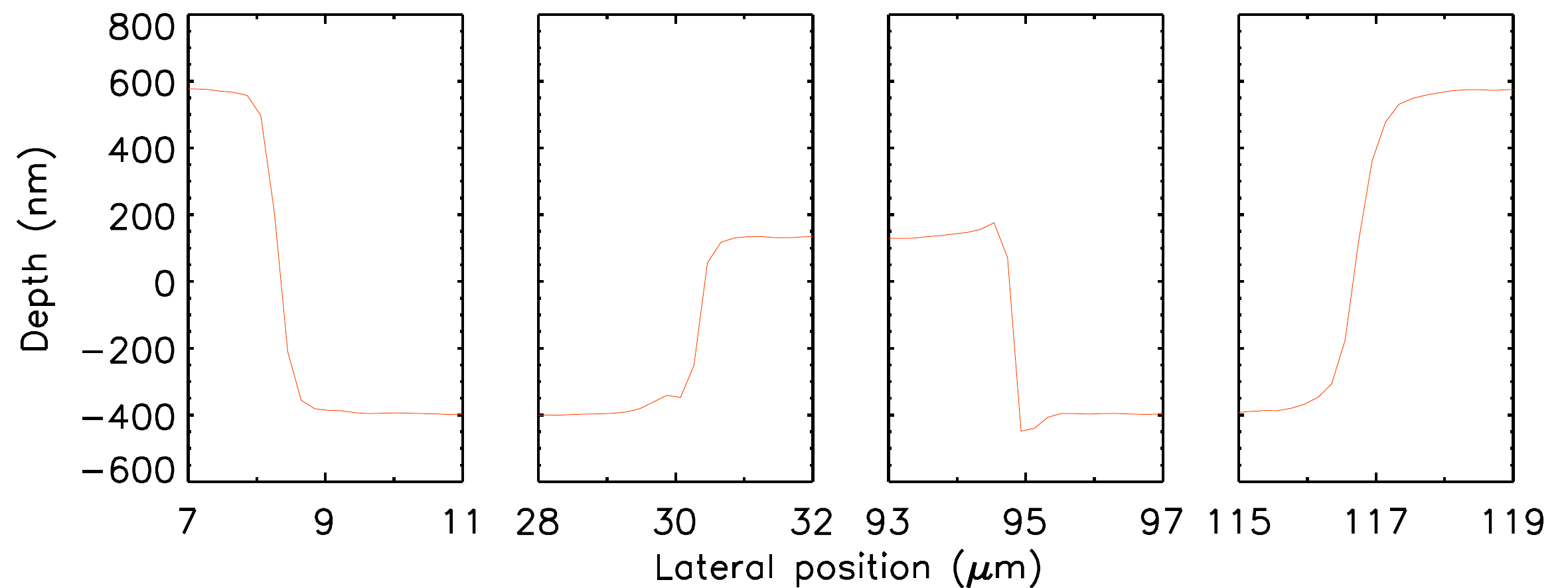}}
\resizebox{\hsize}{!}{\includegraphics{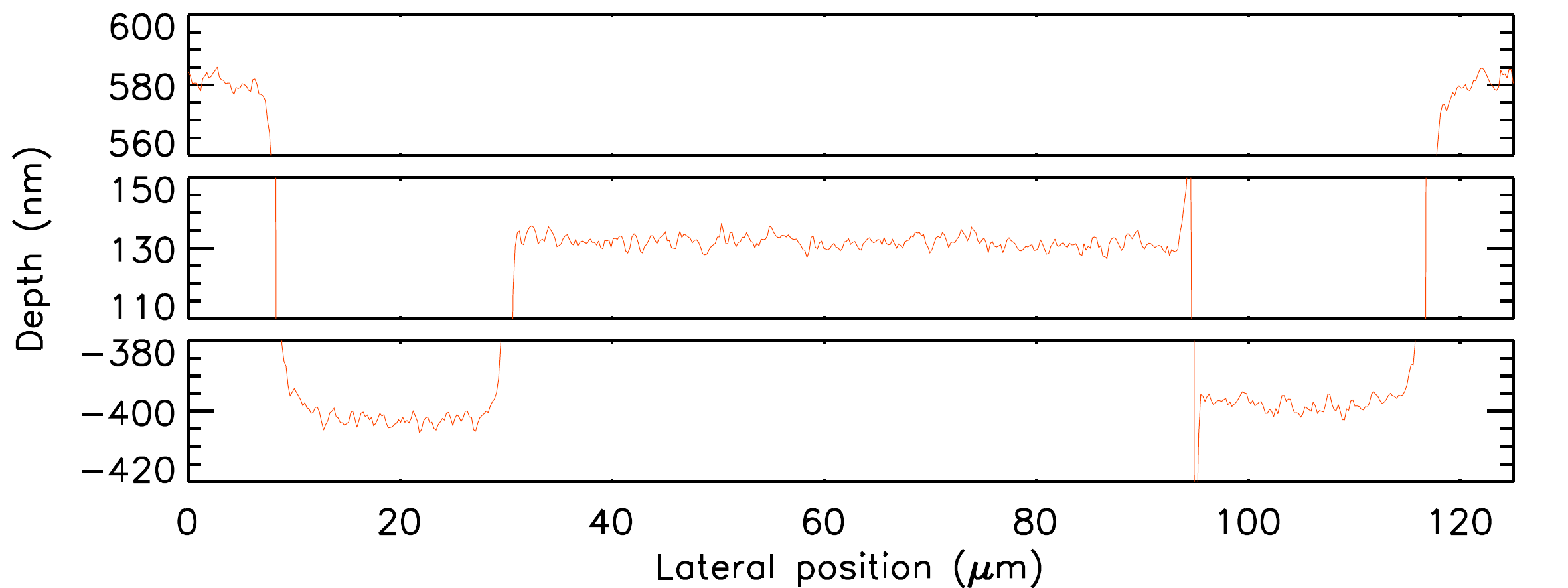}}
        \caption{Profile of the phase mask measured by profilometry. Top: general profile. Middle: horizontal zoom at the mask transitions. Bottom: vertical zoom on the roughness of the different mask steps.}
        \label{Fig_Phase_Mask_Profile}%
\end{figure}

The excellent shape of masks made by this procedure is evident from the optical profilometery made using a Wyko interferential microscope (Fig.\,\ref{Fig_Phase_Mask_Profile}). From this profile we measure the characteristic dimensions of the phase mask defined in Sect.\,\ref{Sec_DZPM}. Values obtained have been included in Tab.\,\ref{Tab_Phase_mask}. The dimensions of the prototype are in good agreement with our specifications showing a relative error better than 1.1\% in all cases (Tab.\,\ref{Tab_Phase_mask}). The root mean square (rms) roughness within the machined areas is 1.8\,nm, identical to that of the substrate outside of the mask. While a slight rounding of the edges can be seen, the transition zone is less than 1\,$\mu$m wide, which is our specifications. 

\begin{table}
\centering
\caption{Dimensions of the central parts of the phase mask specified (specifications) and measured by using a Wyko interferential microscope (measured dimensions). Relative differences between these two sets of parameters are given in percent.} 
\begin{tabular}{|C{1.6cm}|C{1.9cm}|C{1.7cm}|C{1.7cm}|}
\hline
Parameters & Specifications & Measured dimensions & Relative differences\\
\hline
d$_{1}$ & 65.29\,$\mu$m & 64.60\,$\mu$m &    1.1\,\% \\
\hline
Z$_{1}$ & 449.7\,nm     & 446.0\,nm     &    0.8\,\% \\
\hline
d$_{2}$ & 108.4\,$\mu$m & 108.7\,$\mu$m &    0.3\,\% \\
\hline
Z$_{2}$ & 978.1\,nm     & 978.0\,nm     & $<$\,0.1\,\% \\
\hline
\end{tabular}
\label{Tab_Phase_mask}
\end{table}

\section{Coronagraph performance in laboratory} \label{Sec_Lab}
In this section, we present the laboratory performance of the designed DZPM that was obtained on the THD-testbed. We compare the results with numerical simulations that account for most of the laboratory conditions, especially the measured apodization function (Sect.\,\ref{Sec_Setup}) and the measured source spectra.

In Sect.\,\ref{SubSec_IWA} we analyze the sensitivity of the DZPM coronagraph to tip-tilt and determine its inner working angle (IWA). Then in the two following sections, we work with very high contrast in the coronagraphic image by controlling the upstream DM with the SCC (see Sect. \ref{SubSec_THD}) in order to generate a dark hole. In Sect.\,\ref{SubSec_LOA} we measure the DZPM sensitivity to defocus and finally, in Sect.\,\ref{SubSec_Contrast} we characterize its chromatic behavior.

\subsection{Inner working angle and extinction}
\label{SubSec_IWA}

\begin{figure}[h!]
        \centering
        \begin{subfigure}[b]{0.48\textwidth}
        \includegraphics[width = \textwidth]{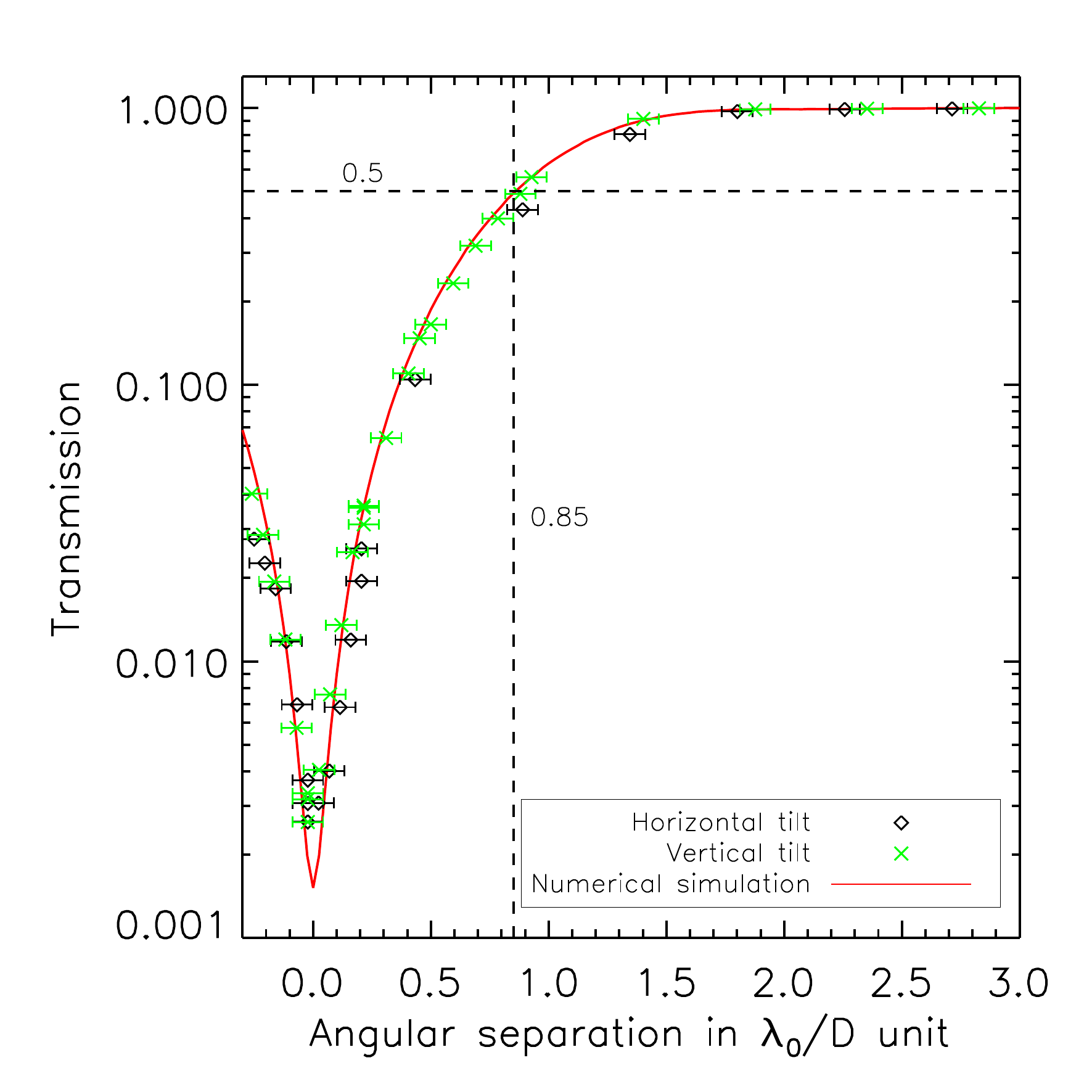}
        \end{subfigure}
        \caption{Intensity transmission of the DZPM as function of tips (black diamonds) and tilts (green crosses) in monochromatic light (637\,nm). The red curve is the function derived from numerical simulations.}
        \label{Fig_IWA}%
\end{figure}

The IWA of a coronagraphic system is defined as the off-axis distance at which the light from a stellar companion is transmitted at 50\%. In this section we determine the IWA of our DZPM prototype in monochromatic light ($\lambda_{0} = 637$ nm). We record series of coronagraphic images for different tips and tilts that are introduced by the tip-tilt mirror upstream from the FPM (Fig.\,\ref{Fig_THD}). Introducing tip-tilt shifts the beam with respect to the FPM and degrades the coronagraphic attenuation. For large shifts, the FPM has no effect and we obtain non-coronagraphic images that are used to normalize the flux in the coronagraphic images. For several tip and tilt, we estimate the transmission by computing the average flux of the normalized image in a disk of $2\,\lambda_{0}/D$ radius centered on the diffraction pattern. Figure\,\ref{Fig_IWA} plots this transmission as a function of tip-tilt. Black diamonds and green crosses represent transmission values obtained by introducing tip and tilt respectively. 
The absolute pointing error has been estimated at $5\:10^{-2}\:\lambda_{0}/D$. This value includes the linearity of the tip-tilt mirror, the drift during the measurement and the precision of the image position measurement. The solid red line represents the theoretical transmission profile obtained by numerical simulations.\\

As expected, tips and tilts have the same impact. Also, the laboratory measurements are consistent with the theoretical curve. We measure an IWA of $0.85\pm0.07\,\lambda_{0}/D$ for our DZPM prototype showing the ability of the concept to observe companions at small angular separations from the star.

From Fig.\,\ref{Fig_IWA}, in an optimal case (without tip and tilt), the DZPM extinction is 500 at 637nm. This extinction is not very high because the prototype is not optimized at 637nm but for a broadband (20\%) between 2 and 10 $\lambda_{opt}/D$ (see Sect.\,\ref{Sec_DZPM}).

\subsection{Focus sensitivity analysis} 
\label{SubSec_LOA}

The DZPM coronagraph theoretically offers its best starlight attenuation over the separations ranging between 2 and 10\,$\lambda_{0}/D$ for a slightly out-of-focus focal plane mask. We here study the sensitivity of our prototype to defocus aberration.

We first put the mask at its optimal position along the optical axis to set the initial conditions for our studies. After first positioning the component at focus, we correct for phase aberrations creating a dark hole by the control of the DM with the SCC. This ensures the right focus positioning of the FPM as the best contrast  obtain after correction. Then, we introduce a defocus aberration on the deformable mirror with different amplitudes and we record two images: one coronagraphic image with the beam centered on the FPM and one non-coronagraphic image with a large tip-tilt. All images are normalized using the non-coronagraphic image at the optimized focus.
\begin{figure}
        \centering
        \begin{subfigure}[b]{0.48\textwidth}
        \includegraphics[width = \textwidth]{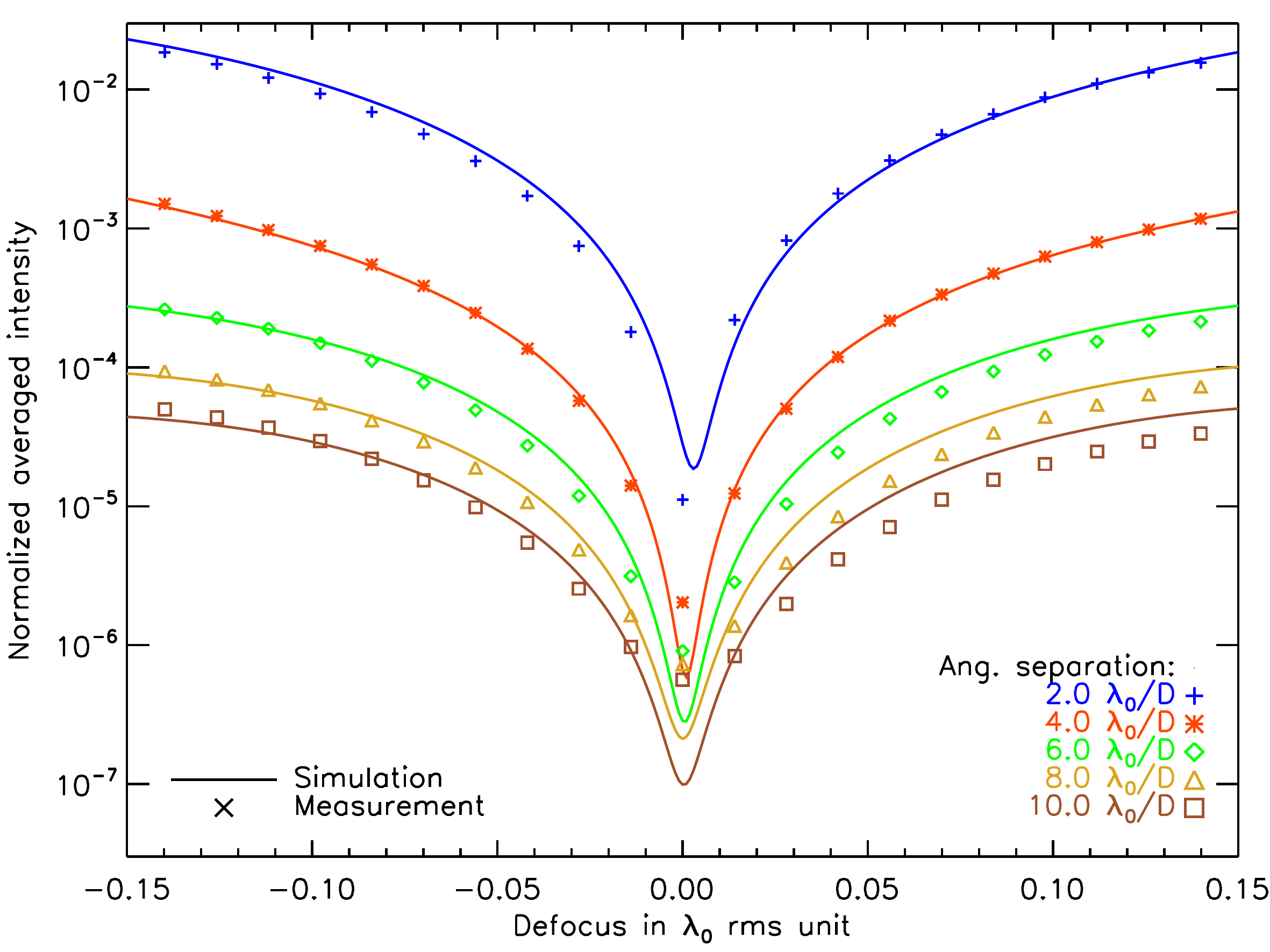}         
        \end{subfigure}
        \caption{Azimuthally averaged intensity of the DZPM coronagraphic images at different angular separations from the star as a function of defocus term in the entrance pupil. Simulated and experimental intensities obtained in monochromatic light (637\,nm) are shown as lines and points respectively. Intensities are averaged over an annulus of 1\,$\lambda_0/D$ width centered at a given separation.}
        \label{Fig_Focus}%
\end{figure}

To estimate the amount of introduced defocus, we compute the encircled energy within the FWHM $I_{fwhm}$ of the non-coronagraphic image at each focus position. Under the regime of small aberrations ($2\pi W \ll 1 \rad$  where $W$ denotes the defocus term expressed with the Zernike coefficient for this mode), we express $I_{fwhm}$ with the Strehl ratio using the Marechal expression ($I_{fwhm}=I_0 \exp(-(2\pi W)^2)$) to derive a theoretical relation between $I_{fwhm}$ and the defocus amplitude. Using this relation, we calibrate the introduced defocus. Then, for each defocus, we measure the average intensity of the coronagraphic image at several angular separations from the central star in $1\,\lambda_{0}/D$ annuli. 

The laboratory results are plotted in Fig.\,\ref{Fig_Focus} and showing good agreement with the theoretical curves. Since we deal with a phase mask coronagraph, contrast degrades quickly as absolute defocus increases as expected. For a defocus of $\lambda_{0}/100$ rms, the contrast is degraded by a factor $\approx \,5$ in laboratory. At the optimized focus position (null defocus), the non-corrected amplitude errors explain the discrepancies between theory and measurements at large separations.

\subsection{Contrast performance in broadband light} 
\label{SubSec_Contrast}

In this section we propose two experiments to probe the performance of the DZPM prototype in terms of contrast and  chromaticity.

\subsubsection{Correction of phase aberrations}
\label{SubSubSec_Attenuation}
\begin{figure}[h!]
        \centering
        \includegraphics[width = 0.24\textwidth]{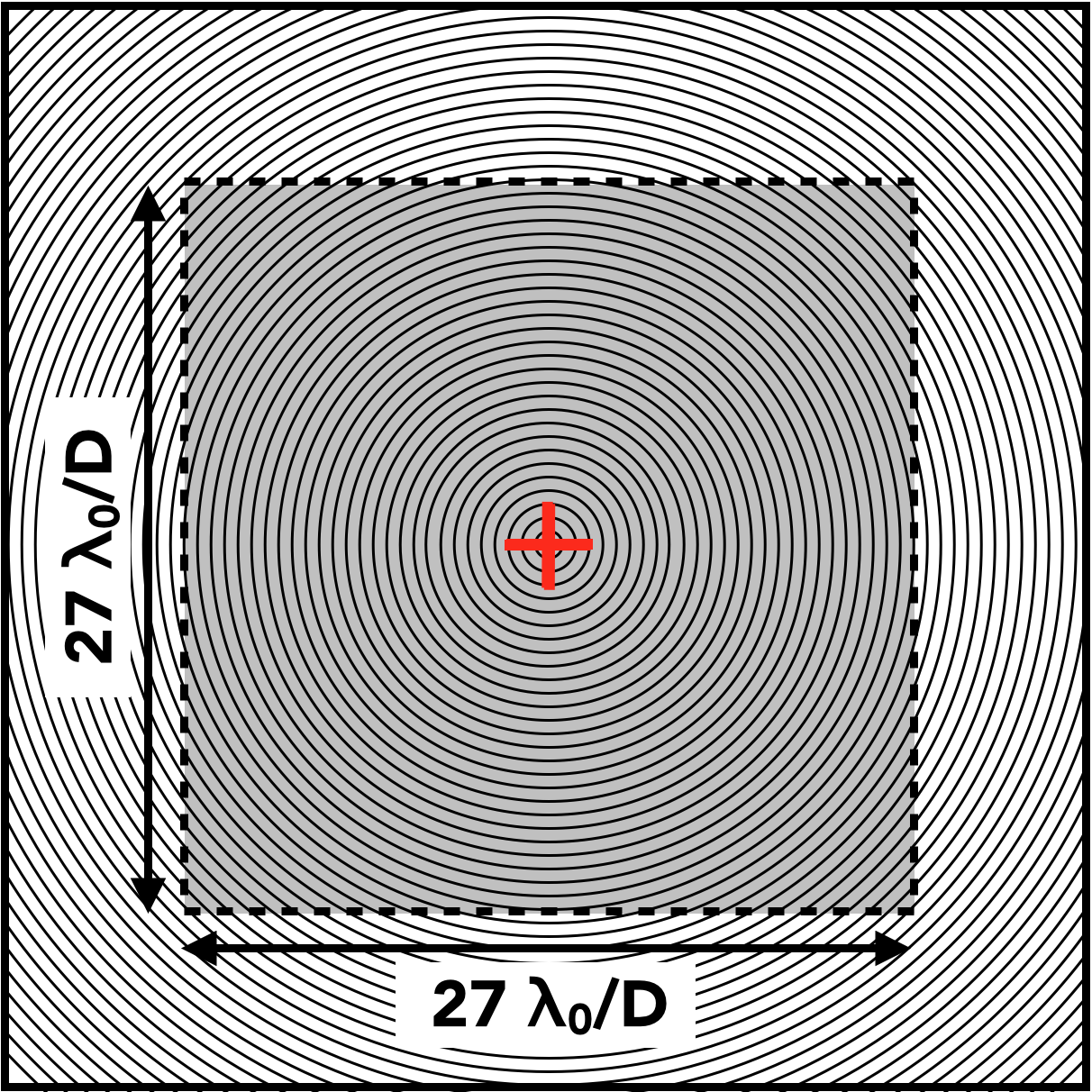}         
        \includegraphics[width = 0.24\textwidth]{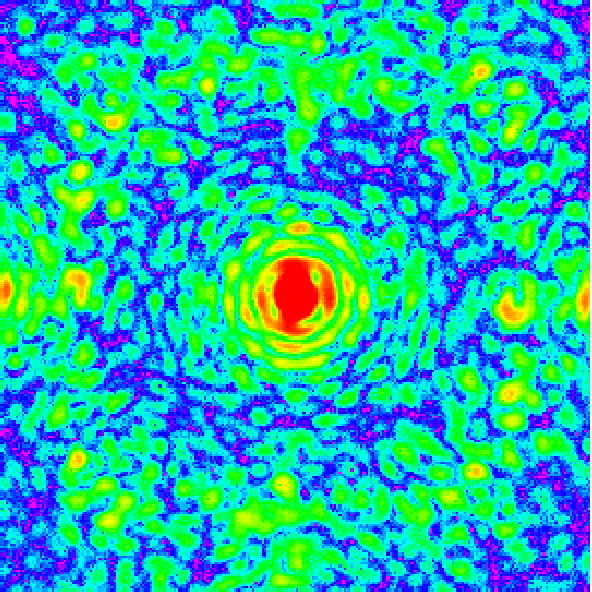}
        
        \includegraphics[width = 0.49\textwidth]{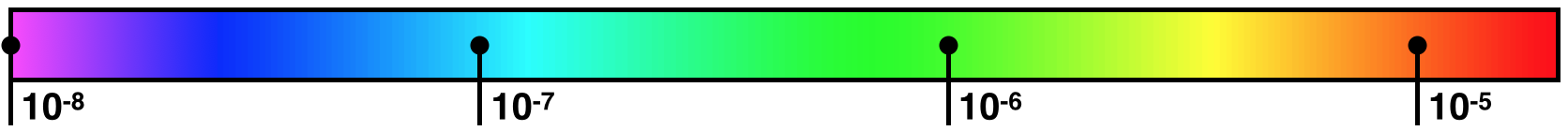}
        \caption{Left: DM influence area (dashed line) and dark hole (gray area). Right: Laboratory DZPM+SCC image obtained in monochromatic light (637\,nm) correcting phase aberrations only. Same spatial scales. Field of view: 40 $\lambda_{0}/D$ by 40 $\lambda_{0}/D$. The color bar associated with this image is the same for all the coronagraphic images in the paper.}
        \label{Fig_Im_exti_mono}%
\end{figure}

First, we attenuated the speckle intensity in monochromatic light inside a dark hole (DH, Fig.\,\ref{Fig_Im_exti_mono} left, gray area) that includes the full DM influence area (Fig.\,\ref{Fig_Im_exti_mono} left, dasched line). Therefore, we were compensating for the phase aberrations only. Fig\,\ref{Fig_Im_exti_mono} (right) shows the image obtained in monochromatic light.

Then, we kept the DM shape frozen and we switched the light source from the monochromatic to the broadband light. For all the 10\,nm and large bandwidths of Tab. \ref{Tab_light_sources}, we recorded two images: one coronagraphic image and one off-axis non-coronagraphic image. The former was normalized by the peak intensity of the latter. Recording the images at all filters with a frozen DM shape ensures that the experimental results are not limited by the SCC performance.
\begin{figure}[h!]
        \centering
        \includegraphics[width = 0.48\textwidth]{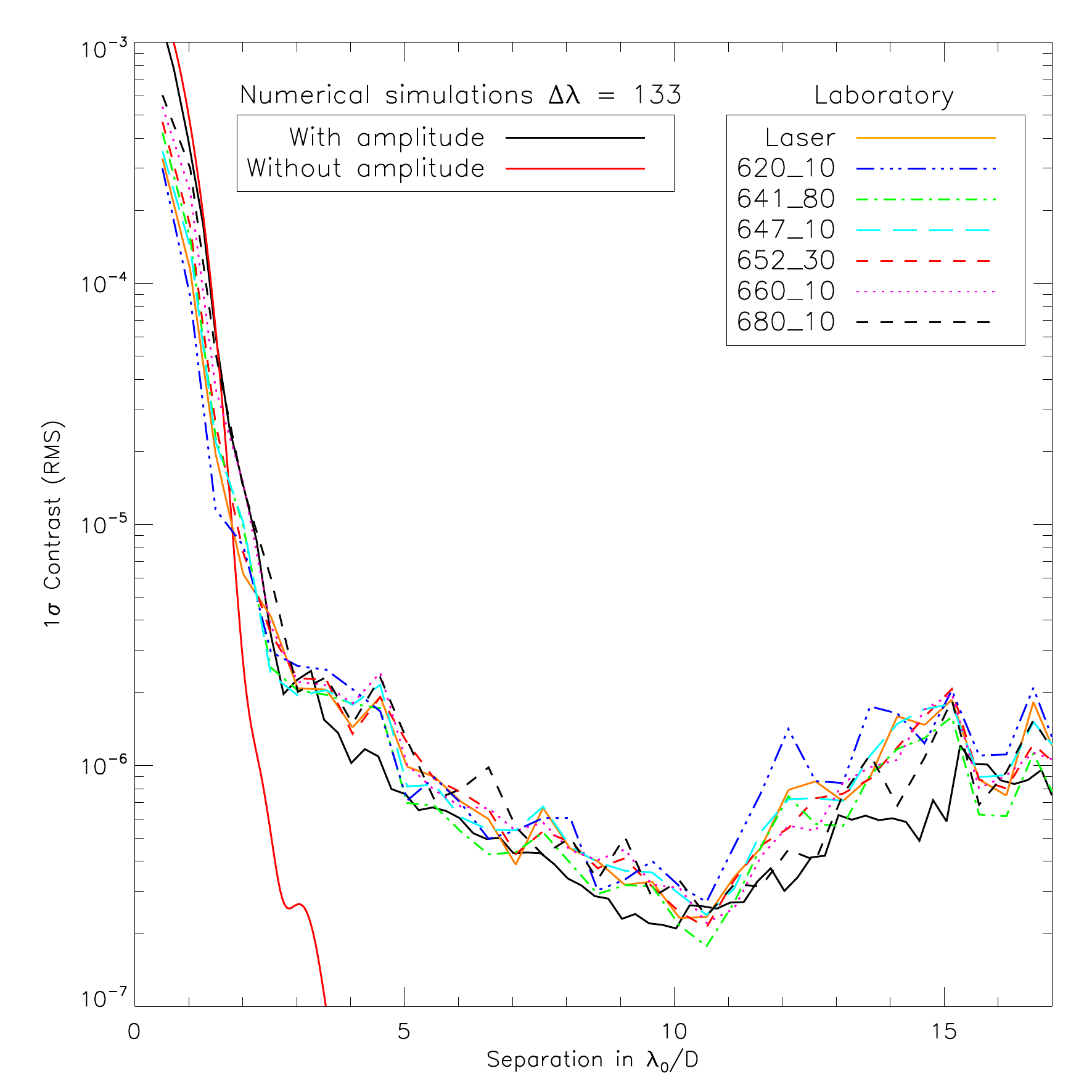}
        \caption{Contrast curves associated with the images obtained in laboratory for the monochromatic light source, the 10 nm bandwidths and the large bandwidths described in Tab.\,\ref{Tab_light_sources}. Two contrast curves derived from numerical simulations for 133\,nm bandwidth are overplotted: with no amplitude aberrations (solid red line) and with amplitude aberrations (solid black line).}
        \label{Fig_DZPM_Lab_vs_simu}%
\end{figure}

For each normalized image, we calculated the azimuthal standard deviation in annuli of width $\lambda_{0}/2D$ centered on the optical axis. The resulting contrast curves plotted in Fig.\,\ref{Fig_DZPM_Lab_vs_simu} reach  $3\,10^{-7}$ for all the filters. This is far from expected performance for the 133\,nm bandwidth without amplitude aberrations (full red line and Sect.\,\ref{SubSec_Optimisation}). However, when the amplitude aberrations measured on the THD-testbed \citep{Mazoyer2013a} are included, the numerical simulations (full black line) are fully consistent with observations. We conclude that the performance obtained in this experiment is limited by amplitude aberrations which cannot be corrected when generating a full dark hole using a single DM.

\subsubsection{Correction of phase and amplitude aberrations} \label{SubSubSec_Contrast}
\begin{figure}[h!]
        \centering
        \includegraphics[width = 0.24\textwidth]{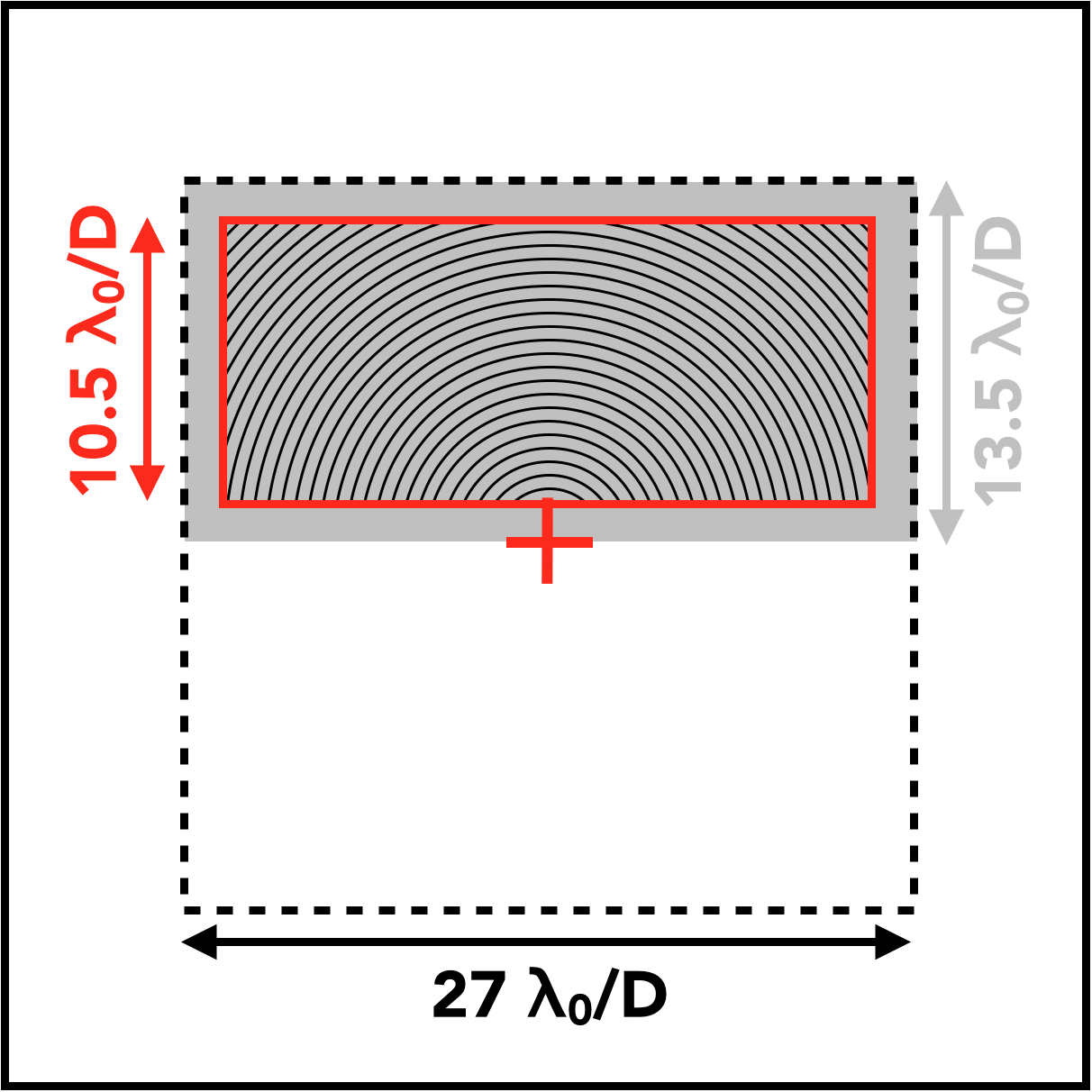}         
        \includegraphics[width = 0.24\textwidth]{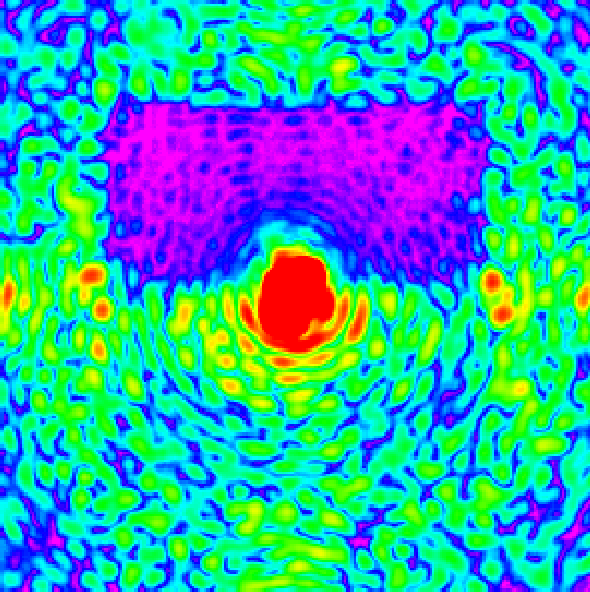}         
        \caption{Left: DM influence area (dashed line), dark hole (gray area) and computation area (full red line). Right: Laboratory DZPM+SCC image obtained in monochromatic light  (637\,nm) correcting phase and amplitude aberrations. Same spatial scale and same colorbar as Fig. \ref{Fig_Im_exti_mono}.}
        \label{Fig_Im_cont_mono}%
\end{figure}

In order to probe the performance of the DZPM prototype, we used the SCC and the DM to minimize the speckle intensity in half of the DM influence area in monochromatic light (see Fig.\,\ref{Fig_Im_cont_mono}), hence correcting the effects of both phase and amplitude aberrations.  Figure\,\ref{Fig_Im_cont_mono} shows the experimental image obtained in monochromatic light (right).

\begin{figure}[h!]
        \centering
        \includegraphics[width = 0.48\textwidth]{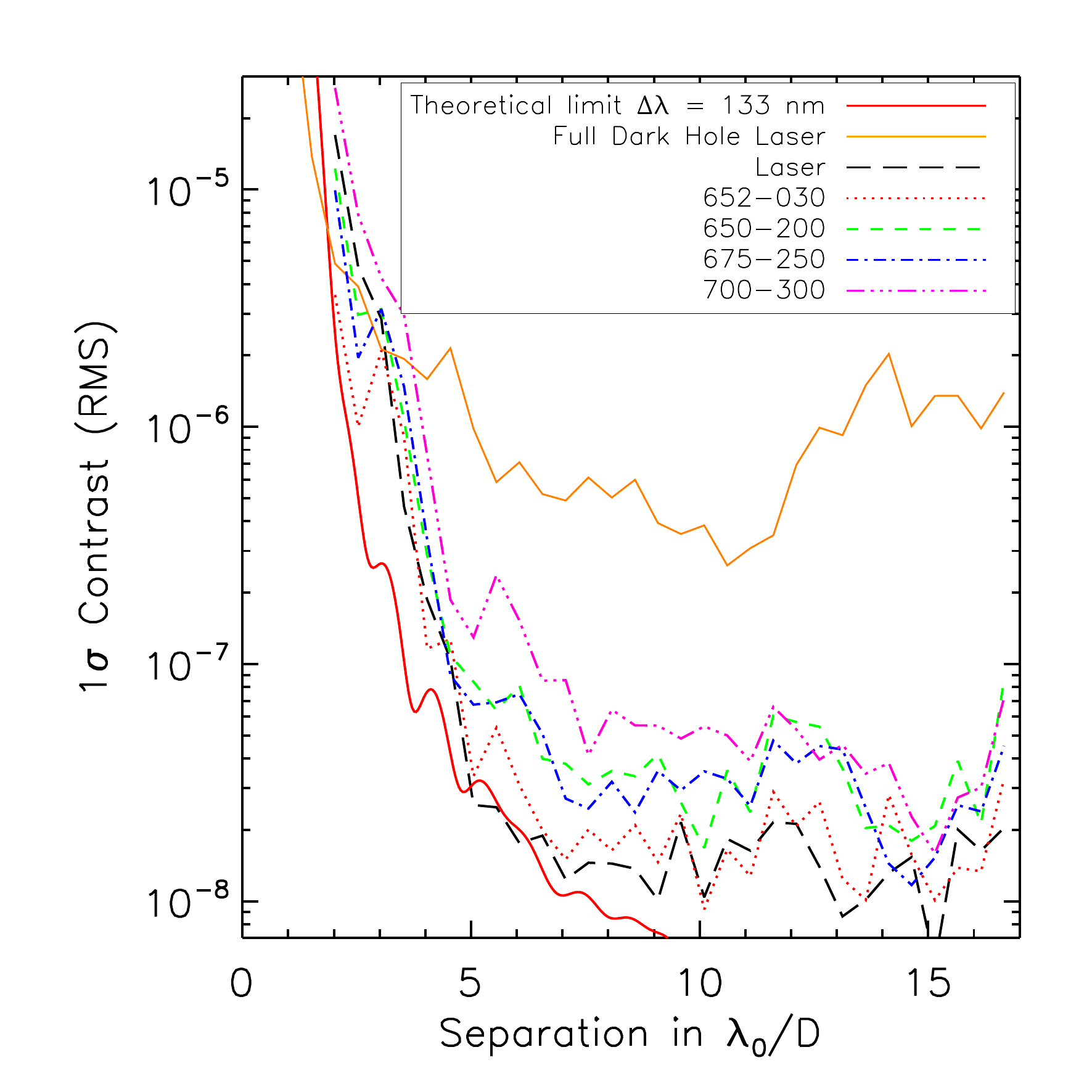}         
        \caption{Contrast curves associated with the laboratory images obtained with a laser source, the 30 nm bandwidth and the very large bandwidths described in Tab.\,\ref{Tab_light_sources}. The experimental performance reached when applying phase-only correction in laser light is overplotted in a cyan dashed curve. The solid red line represents the theoretical limitation derived from numerical simulations with no amplitude aberrations.}
        \label{Fig_Contrast_all_sources}%
\end{figure}

The performance close to the optical axis is limited to a $\approx 500$ extinction (section \ref{SubSec_IWA}). The exposure time is set to detect the residual speckles inside the dark hole. Thus, the center of the image is saturated. The diffraction pattern of the uncorrected speckles located just outside of the DH spread light inside the DH. In order to not bias the contrast estimations, we compute the $1\,\sigma$ contrast curve in a contrast computation area defined as the area within the DH removing 1.5\,$\lambda_{0}/D$ on each side (full red line Fig.\,\ref{Fig_Im_cont_mono} left). This area covers a range of angular separation from 1.5 to $\approx 17\,\lambda_{0}/D$. The $1\,\sigma$ contrast curve associated to the monochromatic image (Fig.\ref{Fig_Im_cont_mono} right) is plotted as black dashed line in Fig.\,\ref{Fig_Contrast_all_sources}. The contrast is enhanced by a factor of $\approx \,40$ with respect to the corrections presented in Sect.\,\ref{SubSubSec_Attenuation} (full orange line). We find that the experimental curve is very close to the expected performance obtained in numerical simulations without amplitude aberrations (full red curve) up to 7\,$\lambda_0/D$. Then, there is a plateau at $<2\,10^{-8}$ between 5 and 17$\,\lambda_0/D$ which corresponds to the limitations of the THD-testbed \citep{Mazoyer2013a}. In conclusion, the prototype reaches the expected performance in monochromatic light.

\begin{figure}[h!]
        \centering
        \includegraphics[width = 0.24\textwidth]{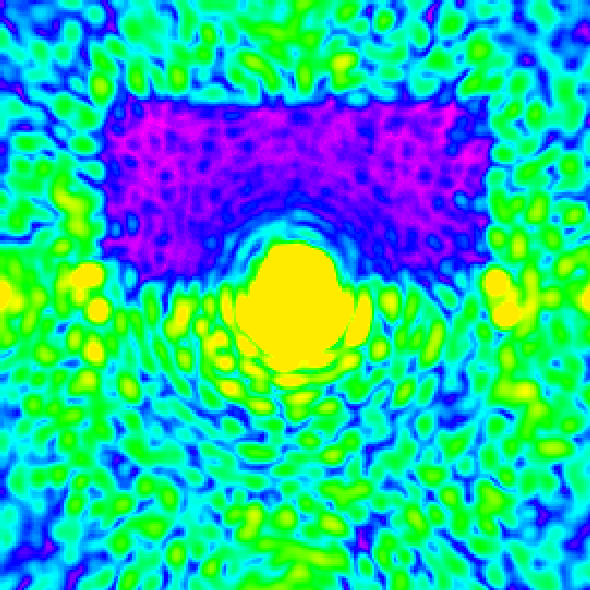}
        \includegraphics[width = 0.24\textwidth]{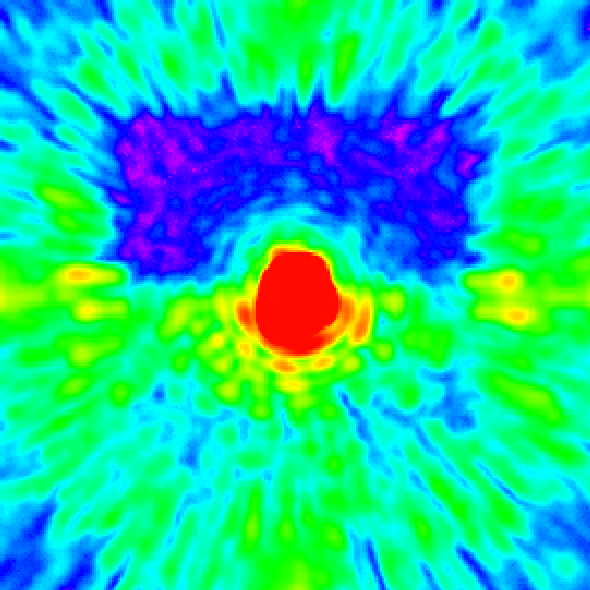}\\
        \vspace{0.06cm}
        \includegraphics[width = 0.24\textwidth]{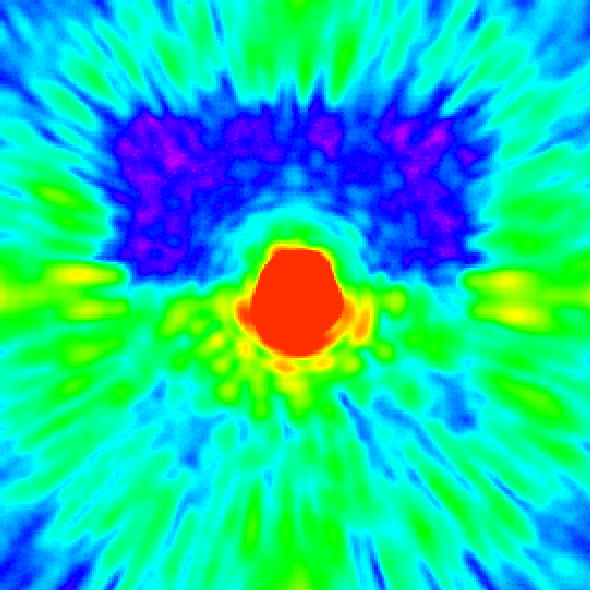}         
        \includegraphics[width = 0.24\textwidth]{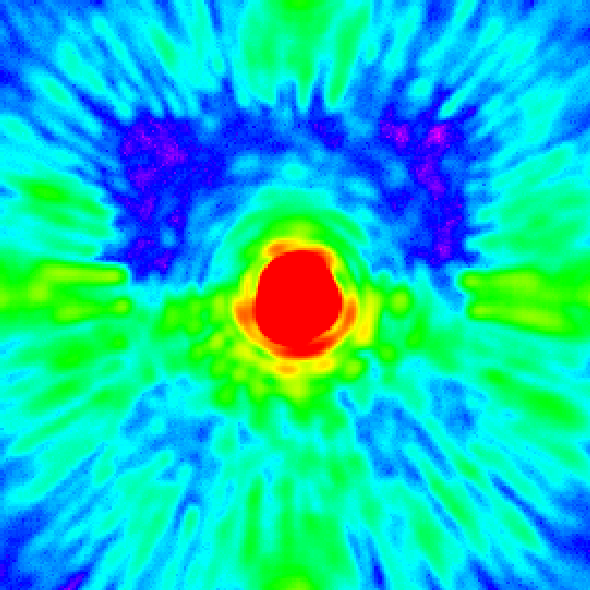}       
        \caption{Laboratory DZPM+SCC images in broadbands: $\Delta\lambda=30\,$nm (top left), $200\,$nm (top right), $250\,$nm (bottom left), and $300\,$nm (bottom right). Same spatial scale and same colorbar as Fig. \ref{Fig_Im_exti_mono}.} 
        \label{Fig_Im_cont_poly}%
\end{figure}

To study the chromatic behaviour of the DZPM coronagraph, we kept the surface shape of the DM frozen and we switched from the laser source to the supercontinuum source using filter-bandwidths from 30\,nm to 300\,nm. The recorded images are presented in Fig.\,\ref{Fig_Im_cont_poly}.

For a bandwidth of 300\,nm, speckles start contaminating the dark hole and they decrease the performance. For $\Delta\lambda=200$ and 250\,nm, the dark hole is not as sharp as for the monochromatic or $\Delta\lambda=30\,$nm case because the speckles are elongated by chromatism. These observations are confirmed when plotting the $1\,\sigma$ contrast curves (Fig.\,\ref{Fig_Contrast_all_sources}).

The monochromatic and the $\Delta\lambda=30\,$nm curves are very similar reaching $<2\,10^{-8}$ levels between $5$ and $17\,\lambda_0/D$. For $\Delta\lambda=200$ and $250\,$nm spectral bandwidths, the performance degrades to $\approx\,7\,10^{-8}$ between $5$ and $7\,\lambda_0/D$ and $\approx\,4\,10^{-8}$ between $7$ and $17\,\lambda_0/D$, which is the expected performance (Sect.\,\ref{SubSubSec_Design}). Finally, we increased the bandwidth to $300$\,nm, corresponding to a bandwidth of $50\%$, and found a degradation of the performance by a factor of 3 only  which is coherent with numerical simulations.
 
All these results demonstrate that the DZPM prototype is chromatically very robust achieving contrast levels better than $7\,10^{-8}$ for a $250$\,nm bandwidth.

\section{Conclusion} \label{Sec_Conclusion}
We performed high-contrast imaging experiments associating a dual-zone phase mask coronagraph and a self-coherent camera in laboratory. We are able to achieve $2\,10^{-8}$ contrast levels between 5 and 17$\,\lambda_0/D$ in monochromatic light ($640$\,nm) and $4\,10^{-8}$ for a bandwidth up to $250$\,nm between 7 and 16$\,\lambda_0/D$. Consistent with the numerical simulations, these results bring conclusive demonstration both of the feasibility of this coronagraph concept and its excellent coronagraphic performance. In particular, we demonstrated capability to achieve the required contrast level and bandwidth for the THD-testbed.

We described the manufacturing and characterization of the two components of the coronagraph, pupil apodizer and focal plane phase mask, and their insertion into the existing THD-testbed. In addition to measuring contrast and bandwidth performance, we also measured the IWA  ($0.85\,\lambda_{0}/D$) and the sensitivity to defocus of our prototype.

The DZPM coronagraph offers several advantages which could be beneficial both for existing instruments and for future generations of high-contrast imaging instruments, ground-based as well as space-based. It is a non-invasive device which can easily be implemented in most existing high-contrast instruments and testbeds, the manufacturing of its components is fully demonstrated and well mastered, and it provides excellent performance for bandwidths up to 40\%. The experimental results presented in this paper were obtained with a design optimized for a circular aperture. However, theoretical studies and numerical simulations have shown that the DZPM coronagraph can also be optimized for arbitrary telescope apertures \citep{Soummer2003a,N'Diaye2010SPIE}, which is encouraging in the context of unfriendly pupils.

Finally, this experimental validation paves the way for the development of novel coronagraph designs, such as the recent solutions proposed by \citet{N'Diaye2016}, that are based on the dual-zone phase mask or more complex masks \citep{Guyon2014}. These masks are manufacturable as of today and should achieve $10^{-10}$ raw contrast at small inner working angle with future segmented aperture telescopes. 

\begin{acknowledgements}
This work was carried out at the Observatoire de Paris (France) and Laboratoire d'Astrophysique de Marseille (LAM) under contract number DA-10083454 with the CNES (Toulouse, France). Moreover, it was partially supported by the National Aeronautics and Space Administration under Grants NNX12AG05G and NNX14AD33G issued through the Astrophysics Research and Analysis (APRA) program. This material is also partially based upon work carried out under subcontract 1496556 with the Jet Propulsion Laboratory funded by NASA and administered by the California Institute of Technology. Finally, we would like to thank L. Pueyo for his support.
\end{acknowledgements}

\bibliographystyle{aa}
\bibliography{report.bib}   

\end{document}